\documentclass[iop, twocolumn]{aastex63}
\usepackage{verbatim,subfigure}
\usepackage{hyperref}
\usepackage{multirow}
\usepackage{orcidlink}
\usepackage{threeparttable}
\usepackage{tabularx}
\usepackage{CJKutf8}
\usepackage{graphicx}
\usepackage{threeparttable}
\hypersetup{hidelinks,
	colorlinks=true,
	allcolors=black,
	pdfstartview=Fit,
	breaklinks=true}
\hypersetup{citecolor=blue,urlcolor=red}

\begin{document}
\begin{CJK}{UTF8}{gbsn}

\title{The Frequency-dependent Modulation Features of PSR J1948+3540}

\correspondingauthor{Na Wang, Feifei Kou; Wenming Yan}
\email{na.wang@xao.ac.cn; koufeifei@xao.as.cn; yanwm@xao.as.cn}

\author{Kaige Chang}
\affiliation{School of Physical Science and Technology, Xinjiang University, Urumqi, Xinjiang 830046, People's Republic of China}
\affiliation{Xinjiang Astronomical Observatory, Chinese Academy of Sciences, Urumqi, Xinjiang 830011, People's Republic of China}

\author{Na Wang}
\email{na.wang@xao.ac.cn(N.Wang)}
\affiliation{Xinjiang Astronomical Observatory, Chinese Academy of Sciences, Urumqi, Xinjiang 830011, People's Republic of China}
\affiliation{Key Laboratory of Radio Astronomy and Technology (Chinese Academy of Sciences), A20 Datun Road, Chaoyang District, Beijing, 100101, P.R.China}
\affiliation{Xinjiang Key Laboratory of Radio Astrophysics, Urumqi 830011, China}

\author{Feifei Kou}
\email{koufeifei@xao.as.cn(F.F. Kou)}
\affiliation{Xinjiang Astronomical Observatory, Chinese Academy of Sciences, Urumqi, Xinjiang 830011, People's Republic of China}
\affiliation{Key Laboratory of Radio Astronomy and Technology (Chinese Academy of Sciences), A20 Datun Road, Chaoyang District, Beijing, 100101, P.R.China}
\affiliation{Xinjiang Key Laboratory of Radio Astrophysics, Urumqi 830011, China}

\author{Wenming Yan}
\email{yanwm@xao.as.cn (W.M. Yan)}
\affiliation{Xinjiang Astronomical Observatory, Chinese Academy of Sciences, Urumqi, Xinjiang 830011, People's Republic of China}
\affiliation{Key Laboratory of Radio Astronomy and Technology (Chinese Academy of Sciences), A20 Datun Road, Chaoyang District, Beijing, 100101, P.R.China}
\affiliation{Xinjiang Key Laboratory of Radio Astrophysics, Urumqi 830011, China}

\author{Jianping Yuan}
\affiliation{Xinjiang Astronomical Observatory, Chinese Academy of Sciences, Urumqi, Xinjiang 830011, People's Republic of China}
\affiliation{Key Laboratory of Radio Astronomy and Technology (Chinese Academy of Sciences), A20 Datun Road, Chaoyang District, Beijing, 100101, P.R.China}
\affiliation{Xinjiang Key Laboratory of Radio Astrophysics, Urumqi 830011, China}

\author{Shijun Dang}
\affiliation{School of Physics and Electronic Science, Guizhou Normal University, Guiyang 550025, Peopleʼs Republic of China}
\affiliation{Guizhou Radio Astronomical Observatory, Guizhou University, Guiyang 550025, Peopleʼs Republic of China}
\affiliation{Guizhou Provincial Key Laboratory of Radio Astronomy and Data Processing, Guiyang 550025, Peopleʼs Republic of China}

\author{Jumei Yao}
\affiliation{Xinjiang Astronomical Observatory, Chinese Academy of Sciences, Urumqi, Xinjiang 830011, People's Republic of China}
\affiliation{Key Laboratory of Radio Astronomy and Technology (Chinese Academy of Sciences), A20 Datun Road, Chaoyang District, Beijing, 100101, P.R.China}
\affiliation{Xinjiang Key Laboratory of Radio Astrophysics, Urumqi 830011, China}

\author{Vishal Gajjar}
\affiliation{University of California, Berkeley, 94720, the United States of America}

\author{Xia Zhou}
\affiliation{Xinjiang Astronomical Observatory, Chinese Academy of Sciences, Urumqi, Xinjiang 830011, People's Republic of China}
\affiliation{Key Laboratory of Radio Astronomy and Technology (Chinese Academy of Sciences), A20 Datun Road, Chaoyang District, Beijing, 100101, P.R.China}
\affiliation{Xinjiang Key Laboratory of Radio Astrophysics, Urumqi 830011, China}

\begin{abstract}

Using observations from GMRT and FAST, we conducted multi-wavelength studies on PSR J1948+3540 and analyzed its intensity modulation characteristics in detail. We found that the intensity modulation of this pulsar exhibits broad low-frequency modulation features. The modulation frequency/period is time-dependent, but the dominant modulation component varies with the observing frequency. Specifically, at low frequencies, the modulation is dominated by the first half of the middle component, while at high frequencies, it is dominated by the second half of the middle component. Spectral analysis revealed that the intensities of the leading and trailing components vary with the observing frequency, but the middle component does not change significantly. Besides, the polarization analyses reveal that the peak of the radiation intensity is located in the latter half of the middle component, whereas the linear polarization is dominant in the former half. However, due to the low degree of linear polarization, the change of the dominant modulation component with the observed frequency is not caused by the variation in linear polarization. The phenomenon of the dominant modulation component varying with observing frequency has not been reported before and remains difficult to understand within the current theoretical framework.

\end{abstract}

\keywords{pulsars: general$-$ stars: neutron$-$ pulsars: individual (PSR J$1948+3540$)}

\section{INDRUCTION}

Pulsars are known for their stable radiation characteristics, including their consistent rotation periods and integrated average profiles. Since their discovery, extensive research, analysis, and statistical work have been conducted on pulsar radiation characteristics, aiming to gain a more comprehensive understanding of pulsar radiation mechanisms and magnetospheric structures. Observations have revealed that radiation intensity and profiles of some pulsars vary with observing frequency.
By analyzing observational data at different frequencies, it is found that the radiation intensity of pulsars typically follows a power-law relationship, indicating that the radiation intensity of pulsars gradually decreases as the frequency increases  \citep{Malofeev_1980,Rankin_1983a,Rankin_1983b,2000A&AS..147..195M,2018MNRAS.473.4436J}.
 Detailed analysis also revealed that the spectrum varies in the different parts (or components) of the profile \citep{Rankin_1983a,Rankin_1983b,Lyne_1988, Kramer_1994,2007_Chen,2021ApJ...917...48B, 2022ApJ...927..208B}. Additionally, the radiation radius of pulsars also exhibits a power-law relationship with frequency, resulting in an increase in radiation radius with decreasing frequency, known as radius-to-frequency mapping (RFM for short) \citep{Komesaroff_1970, Cordes_1978}. 

A typical phenomenon in pulsar radiation is the modulation of intensity, manifesting as significant variations in the radiation intensity of the entire or partial components over observation time, often with quasi-periodicity.
A classic example of this phenomenon is sub-pulse drifting, typically observed in the conal components of the pulse profile. 
This drifting phenomenon may be linked to particle beams in the vacuum gap over the polar cap rotating around the magnetic axis due to the $\vec{E}\times\vec{B}$ drift of aligned or anti-aligned rotators \citep{Drake_1968,Backer_1970,Ruderman_1975}, or to variable  $\vec{E}\times\vec{B}$ drift of partially screened gap (PSG) sparks lagging the pulsar's corotation velocity \citep{2020MNRAS.496..465B,2022_Basu}.
The two typical periodic modulations, in addition to sub-pulse drifting, are periodic nulling and periodic amplitude modulation, which are seen in both the core and conal components simultaneously \citep{2007MNRAS.380..430H,2009MNRAS.393.1391H,2017ApJ...846..109B}.
Although the physical origins of periodic nulling and periodic amplitude modulation remain unclear, it is generally accepted that they are distinct from sub-pulse drifting \citep{Basu_2020}. Both sub-pulse drifting and periodic amplitude modulation are variations in time series. The time-dependent modulation are common, but the frequency dependent modulation are rarely reported, which are great challenge to the understanding of pulsar radiation and magnetosphere structure.

PSR J$1948+3540$ (B$1946+35$) is a radio pulsar discovered by the Jodrell Bank Mk I telescope at $408\, \mathrm{MHz}$, with a period of $0.717(s)$ \citep{Davies_1970,Hobbs_2004}. Based on its integrated pulse profile and observational properties, PSR J$1948+3540$ was categorized as ``St'', which signifies having a core-single profile \citep{Rankin_1993a,Rankin_1993b}. By fitting the widths of the pulsar profiles at various frequencies, the emission geometry could be modeled by calculating the core-component width at $1 \, \rm GHz$, and a inclination angle of $32^\circ$ was given for this pulsar \citep{mitra2017MNRAS.468.4601M}. 
 According to previous analysis, PSR J$1948+3540$ exhibits intensity modulation, but the period of the intensity modulation varies with observation time \citep{mitra2017MNRAS.468.4601M}. 
It is suggested that the intensity modulation period/frequency is time-dependent rather than frequency-dependent.
Based on the multi-wavelength observational data analysis using the GMRT and FAST telescopes, we found that the modulation characteristics are not only time-dependent but also frequency-dependent. In this paper, to clearly present the emission properties of PSR J1948+3540, we utilized the largest single-dish telescope FAST and the aperture synthesis radio telescope GMRT to make the multi-band observations for this pulsar. The high sensitive multi-band observations of PSR J$1948+3540$ reveals previously unknown and more complex emission properties, which may further complicate the understanding of radio emission in this pulsar.  The observations and the data processing method
are introduced in Section \ref{section:2}. The results are given in Section \ref{section:3}. We
summarize our results and discuss the possible geometric structure of the magnetosphere in Section \ref{section:4}.

\section{OBSERVATIONS AND DATA PROCESSING}\label{section:2}

The Five hundred-meter Aperture Spherical radio Telescope (FAST for short)  is located at Guizhou, China, with an illuminated aperture of $300\, \, \mathrm{m}$ in operation. Since July $2018$, a $19$-beam receiver, designed to cover the frequency ranges from $1.05$ to $1.45\, \, \mathrm{GHz}$, has been installed and operational. The gathered data is processed by a digital backend system utilizing the Reconfigurable Open-architecture Computing Hardware, version$2$ (ROACH2), as described by \citet{2019_JP_FAST,2020_Jiang_19beam}. These data are then stored in the PSRFITS data format specifically tailored for search-mode operations \citep{Hotan2004PASA...21..302H},  with a time resolution of $49.152$ microseconds and a frequency resolution of $0.122\, \, \mathrm{MHz}$. We conducted a $5$-minute of polarization calibration and $45$-minute observation of PSR J$1948+3540$ using FAST on October $5$, $2020$ (MJD $59127$), and got a total of $3765$ pulses.  In the FAST data processing, individual pulses were extracted using DSPSR \citep{van2011PASA...28....1V} according to the ephemeris provided by PSRCAT \citep{Manchester2005AJ....129.1993M}.
Then the band edge in the data were eliminated by Pulsar Archive Zapper (PAZ) plug-in of PSRCHIVE \citep{Hotan2004PASA...21..302H}. 
And the narrow-band and impulsive radio-frequency interference were flagged and removed by PAZI and PAZ plug-in of PSRCHIVE.
During polarization calibration, Stokes parameters are acquired by the pac package of PSRCHIVE after calibrating the observations using the folded calibration file. Subsequently, TEMPO2 \citep{Hobbs2006MNRAS.369..655H} was used to measured Dispersion measure (DM), and rmfit was used to measured Rotation measure (RM), with both corrections applied to the ephemeris as necessary. 

The upgraded Giant Metrewave Radio Telescope (uGMRT), located in India, is a synthesis aperture telescope used for radio astronomy. It consists of thirty antennas, each with a diameter of $45\, \, \mathrm{m}$, and operates within the low-frequency range of $110$ to $1460\, \, \mathrm{MHz}$.
Our data were taken on June $2$, $2024$ (MJD $60463$) at  Band $3$ (from $300 \, \rm MHz$ to $500\, \rm MHz$) and June $8$, $2024$ (MJD $60469$) at  Band $4$ (from $550 \, \rm MHz$ to $950\, \rm MHz$) with $3$ hours for each. The raw data  were converted into filterbank files using ugmrt2fil (\url{https://github.com/inpta/ugmrt2fil}). After removing the interference in both the time domain and the frequency domain, we obtained $10101$ and $5942$ consecutive pulses at  band $3$ and  band $4$ respetively. Due to the lack of polarization information, the following analysis of the GMRT data will only focus on its total intensity. The detailed observational information is listed in Table \ref{tab:observation}.

\begin{table*}[]
\begin{threeparttable} 
\centering
\caption{Details of the observational parameters and the analysis results}
      \begin{tabular}{c c c c c c c c c }
        \hline
        Telescope & Center Frequency & MJD  & BW & Pulses & $P_3$ & &Phase delay & \\
        
        &  &   &  &  &  & $\rm{Com_L}\&\rm{M_{II}}$ & $\rm{M_{I}}\&\rm{M_{II}}$ & $\rm{com_{T}}\&\rm{M_{II}}$\\
        
        & (MHz) & & (MHz) & & (cpp) &  & ($\circ$) &  \\
        \hline
         FAST &1250 & 59127  & 400 & 3765 &$0.018(3)$ & $0^{\circ}$ or $-60^{\circ}$ & /&$-160^{\circ}$\\
         & & & & &   $0.023(5)$ & $-30^{\circ}$ & / &$-180^{\circ}$\\
         & & & & &   $0.012(4)$ & $-30^{\circ}$ & / &$-160^{\circ}$\\
         GMRT & 750 & 60469 & 400 & 5942 & $ 0.012(2)$ & $-60^{\circ}$ & $90^{\circ}$ & / \\
         & & & & &   $0.021(3)$ & $-80^{\circ}$ & $60^{\circ}$ & $-180^{\circ}$\\
         & & & & &   $0.018(3)$ & $-150^{\circ}$ & $50^{\circ}$ & 110 \\
         GMRT & 400 & 60463 & 200 & 10101 & $ 0.0058(2)$ & / & / & / \\
         & & & & &   $0.012(3)$ & / & /& /\\
         & & & & &   $0.017(3)$ & / & / & /\\
         \hline
    \end{tabular}
    \label{tab:observation}
      \begin{tablenotes} 
		\item Notes:The phase delay is only an approximate estimate, intended to demonstrate that these radiation components are phase-locked rather than varying in perfect synchrony. At 750~MHz and 1250~MHz, the symbol ``/'' indicates that the distribution is too scattered to yield a meaningful value. At 400~MHz, due to scattering effects and the weaker leading component emission, we did not perform profile component separation. Thus, we only report the modulation period to confirm that these modulated features persist across all frequency bands and epochs of observation.
     \end{tablenotes} 
\end{threeparttable} 
\end{table*}

\section{RESULTS} \label{section:3}
  
\begin{figure*}
    \centering
    \includegraphics[width=0.3\linewidth]{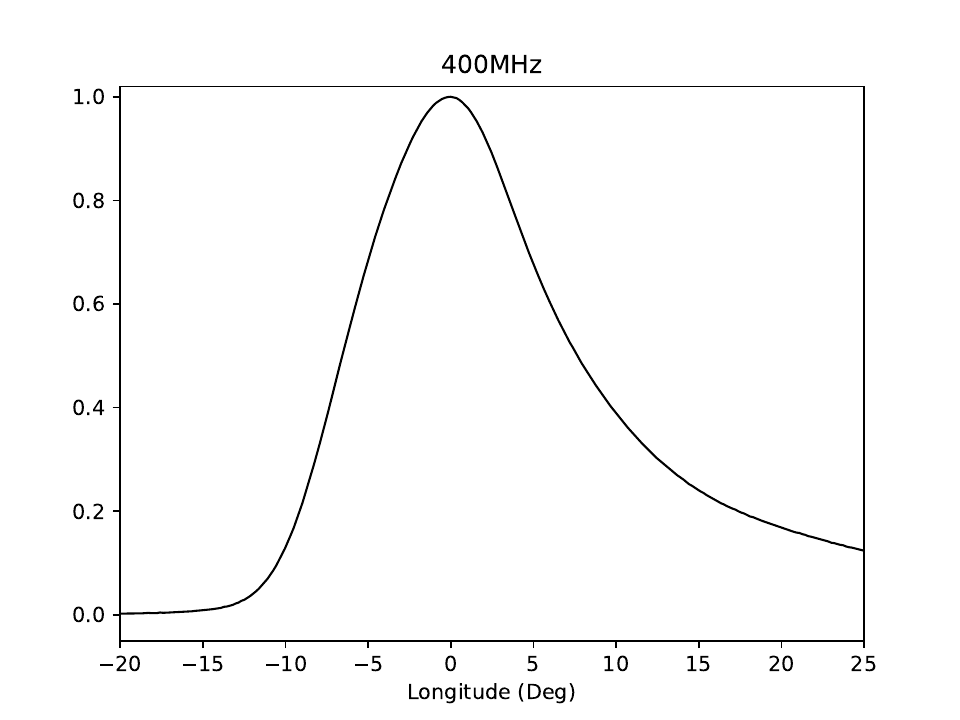}
    \includegraphics[width=0.3\linewidth]{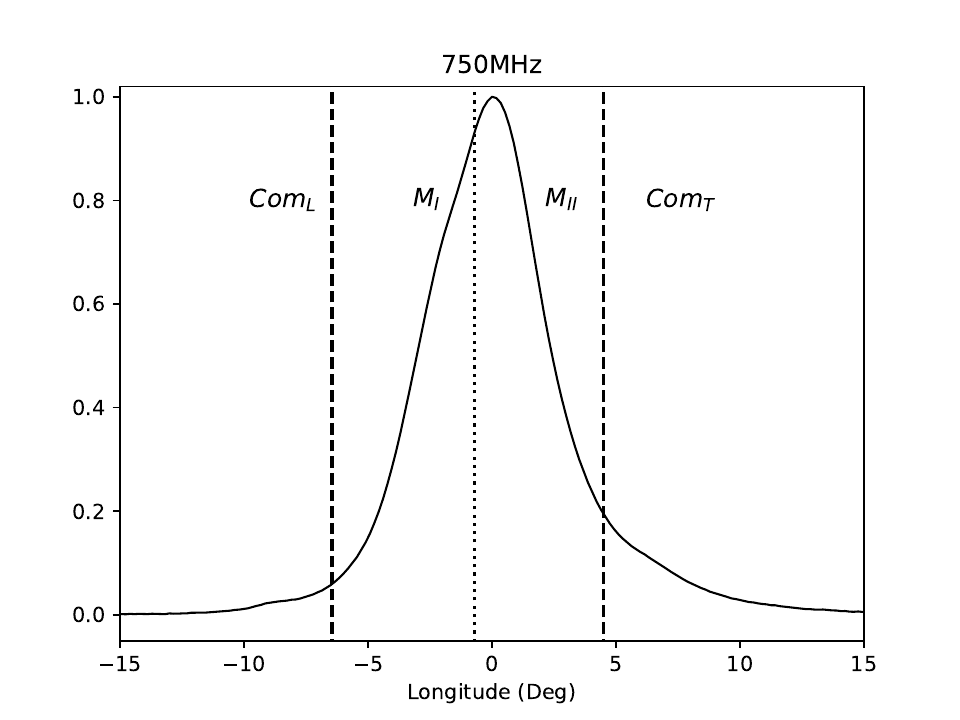}
    \includegraphics[width=0.3\linewidth]{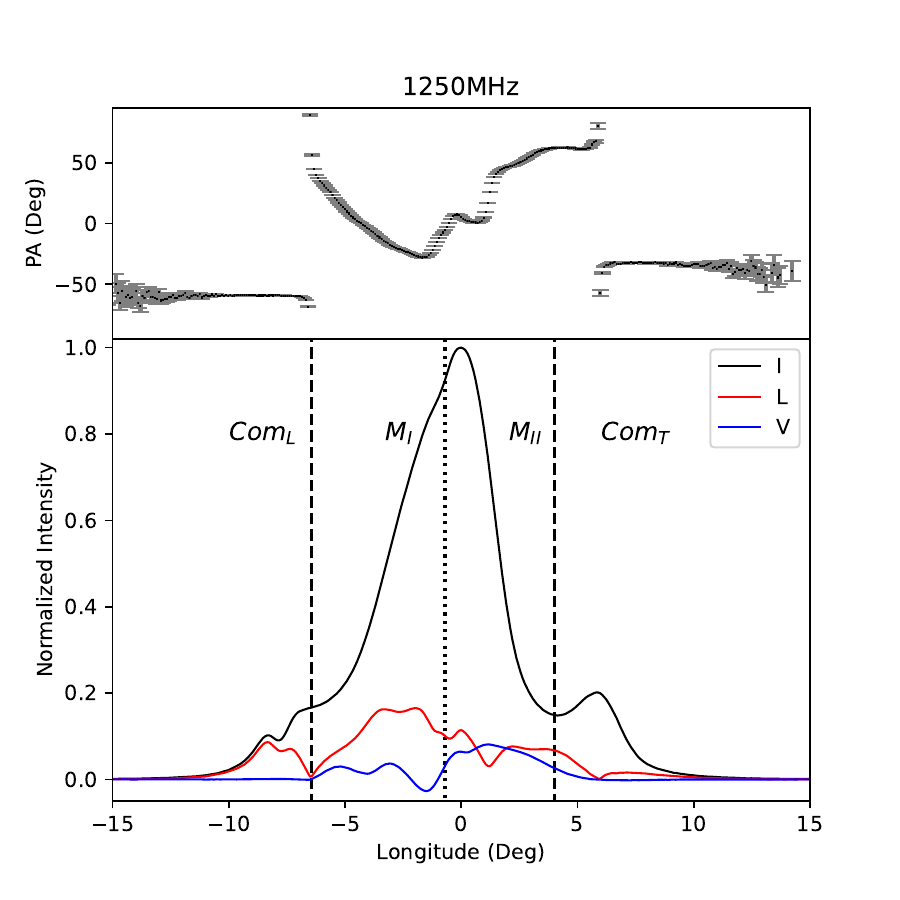}
    \caption{The normalized pulse profile for PSR J1948+3540 at center frequencies of $400\, \rm MHz$, $750\, \rm MHz$ and $1250\, \, \mathrm{MHz}$. The longitude of the pulse peak is set to be zero. The top panel of the right column shows the the position angles of the linearly polarized emission. The black, red, and blue lines of the bottom panel are the total intensity, linear polarized intensity, and circular polarized intensity, respectively. We distinguish the pulse phases of distinct components using black dashed and dotted lines. Here, $\rm{Com_{L}}$ (leading component) and $\rm{Com_{T}}$ (trailing component) denote the conal component, while $\rm{M_{I}}$ and $\rm{M_{II}}$ are the first half and the second half of the middle(core) component, respectively.}
    \label{fig:prof}
\end{figure*}

To get the integrated profiles, the GMRT and FAST data are respecitvely centered to $400\, \rm MHz$, $750\, \rm MHz$ and $1250\, \, \mathrm{MHz}$. The normalized integrated profiles are shown in Figure~\ref{fig:prof}, with the FAST data showing the polarization profiles. The polarization degree of this pulsar is relatively low, with a linear polarization degree of only $20\%$ and a circular polarization degree of only $5\%$ at L band. 

The pulse profile of this pulsar exhibits a complex structure, comprising a central core emission component (hereafter referred to as the middle component) flanked by two relatively weak conal components (designated as the leading and trailing components). The integrated profiles of PSR J1948+3540 at 1.4~GHz and 4.6~GHz are presented in the paper of \citet{mitra2017MNRAS.468.4601M}, showing significant evolution in the intensities of the leading and trailing components with frequency, especially at 4.6 GHz where the intensities of the leading and trailing components exceed that of the middle component. Besides, as reported by \citet{mitra2017MNRAS.468.4601M}, the core component itself comprised of two overlapping Gaussian-like structures. For clarity in subsequent descriptions, we designated the two conal components as $\rm{Com_{L}}$ (leading component) and $\rm{Com_{T}}$ (trailing component). While for the middle (core) component, we had further divided it into the first half and the second half based on the total intensity profile, naming them $\rm{M_{I}}$ and $\rm{M_{II}}$ respectively. (Note: This classification is purely for descriptive convenience in analyzing phenomenological features and does not imply distinct physical origins of the emission components.)

From figure~\ref{fig:prof}, the observational results at 750~MHz and 1250~MHz reveal a significant evolution of the integrated profile with frequency, manifested as an increase in the intensities of the leading and trailing components ($\rm{Com_{L}}$ and $\rm{Com_{T}}$), and more details are presented in Figure \ref{fig:delx}. However, at 400 MHz, the pulsar profile is affected by significant scattering effects. Hence, no component separation was attempted at this frequency.
Our subsequent analysis results are primarily based on the data obtained at 750~MHz and 1250~MHz, although scattering broadening at 400~MHz does not affect the conclusions of the article.

\subsection{The single pulse modulation}
\begin{figure*}
    \centering
    \includegraphics[width=0.3\linewidth]{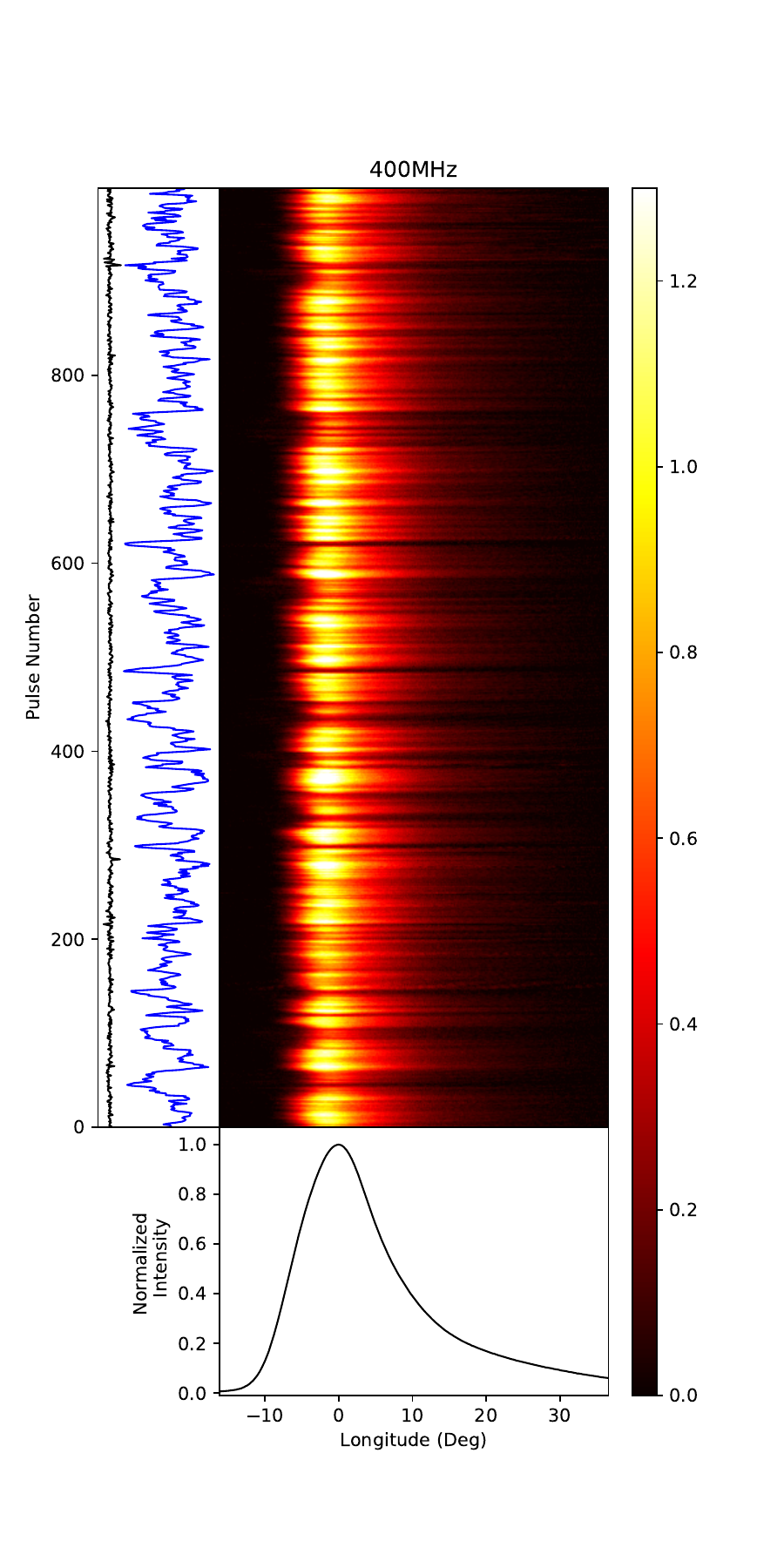}
    \includegraphics[width=0.3\linewidth]{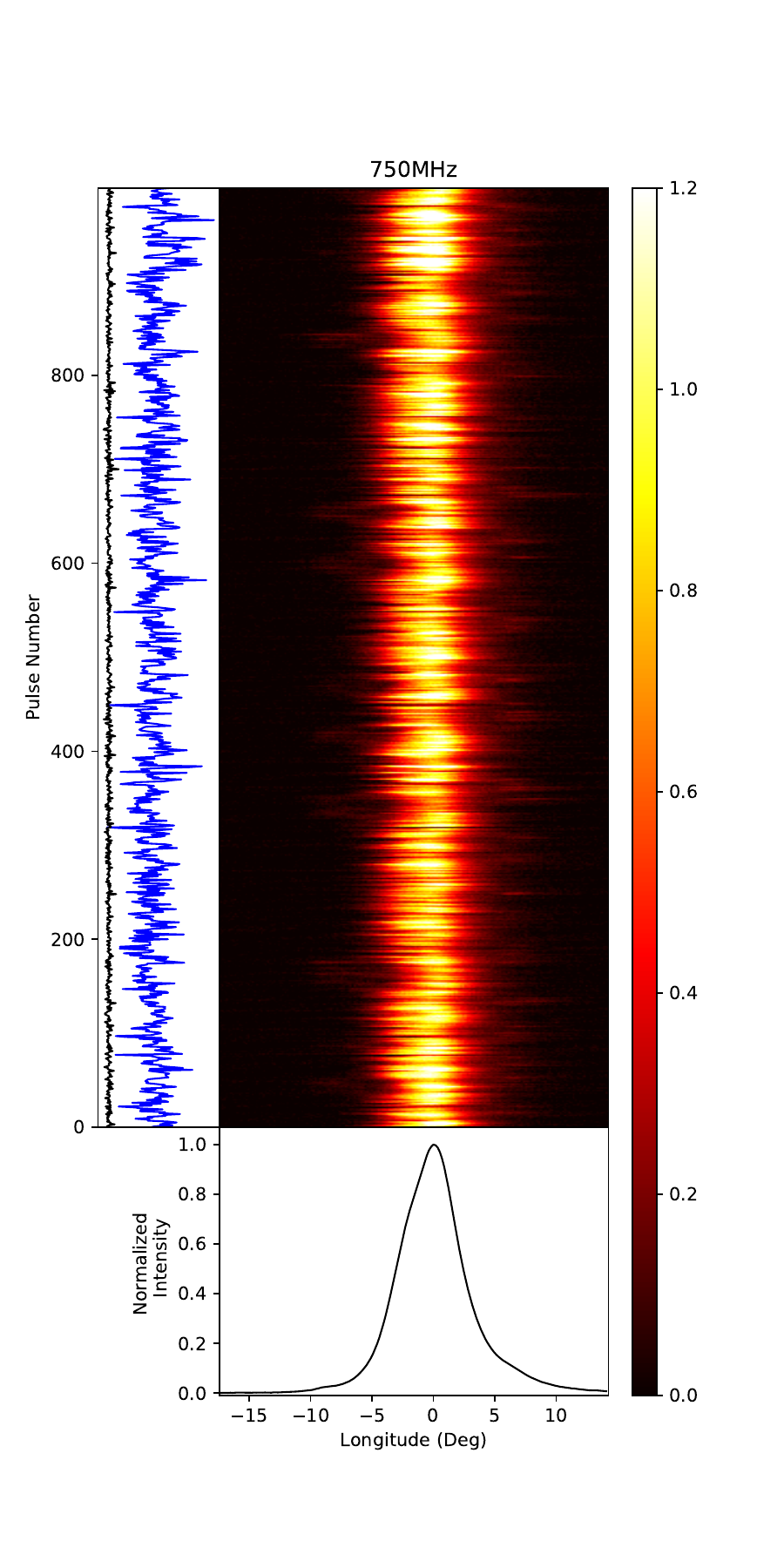}
    \includegraphics[width=0.3\linewidth]{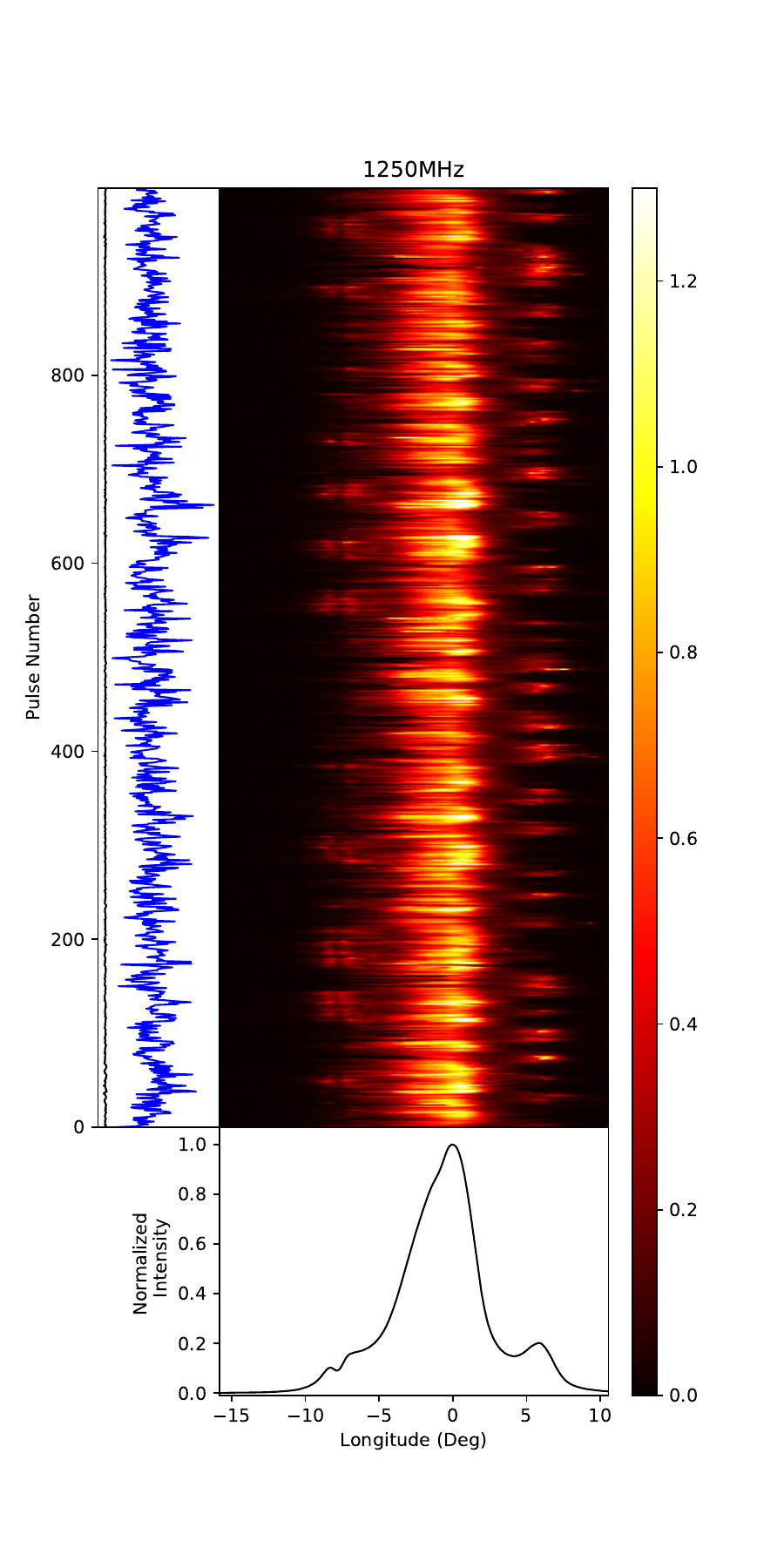}
    
    \includegraphics[width=0.45\linewidth]{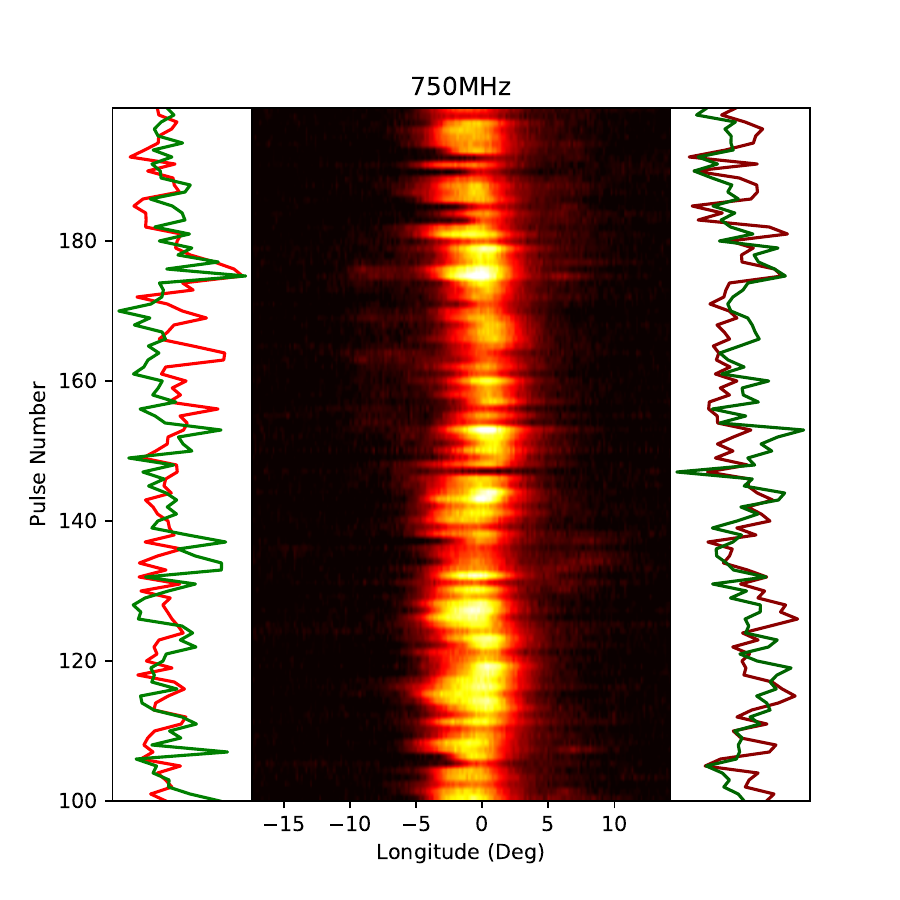}
    \includegraphics[width=0.45\linewidth]{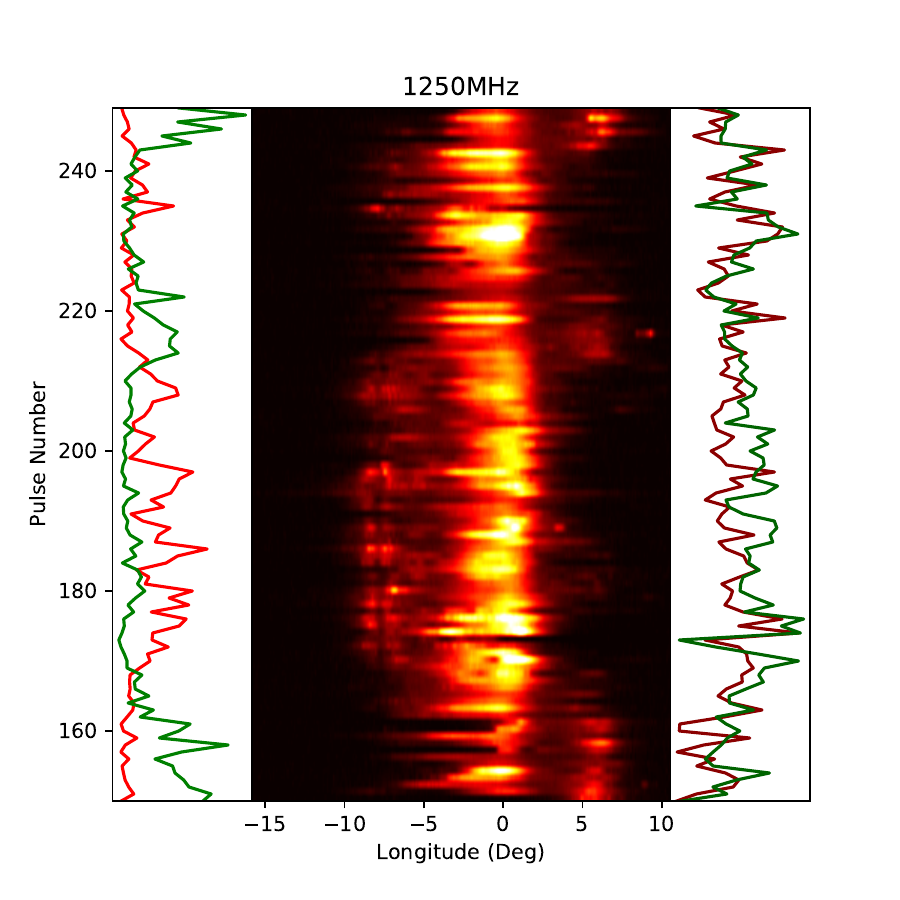}
    \caption{The single-pulse stacks of PSR J1948+3540 at center frequencies of $400\, \, \mathrm{MHz}$ (upper left), $750\, \, \mathrm{MHz}$ (upper middle) and $1250\, \, \mathrm{MHz}$ (upper right). The left panels of top columns show the energy variations for the on-pulse range (blue solid line) and the off pulse range (black solid line). The bottom panels of top columns show the integrated pulse profiles which are normalized to the peak intensity. The two columns below display enlarged images for 750~MHz and 1250~MHz, with pulse sequences spanning from 100-200 and 150-250, respectively. The left side shows the energy of $\rm{Com_{L}}$ (red) and $\rm{Com_{T}}$ (green), while the right side depicts the energies of $\rm{M_{I}}$ (dark red) and $\rm{M_{II}}$ (dark green), respectively.}
    \label{fig:stack}
\end{figure*}

The single-pulse stacks of PSR J1948+3540 at center frequencies of $400\, \, \mathrm{MHz}$, $750\, \, \mathrm{MHz}$ and $1250\, \, \mathrm{MHz}$ are shown in Figure \ref{fig:stack}. As shown in the figure, the main component exhibits significant intensity variations across the three bands. At $750 \, \rm MHz$ and $1250\, \rm MHz$, we can see that the leading and trailing component undergoes a transition from null to radiation. Moreover, it seems that the emission of the leading and trailing components fluctuates alternately, means that they are phase-locked.  To present these phenomena more visually, enlarged diagrams with a short pulse sequences are shown in the bottom of Figure \ref{fig:stack} : the pulse ranges from 100 to 200 at 750~MHz and from 150 to 250 at 1250~MHz. As shown in the bottom sub-plots, the intensities of these four components all exhibit significant intensity fluctuations. In particular, there are alternating variations in the energy levels of $\rm{Com_{L}}$ and $\rm{Com_{T}}$ (from strong to weak and vice versa), these variations are not fully synchronized, indicating that there is a time delay in their intensity modulation. This phenomenon will be discussed in detail in the following section. At $400 \, \rm MHz$, due to the scattering effect as well as the weak radiation of the leading and trailing components, it is difficult to distinguish them in phase and no significant intensity variations was detected.

\begin{figure*}
    \centering
    \includegraphics[width=0.32\linewidth]{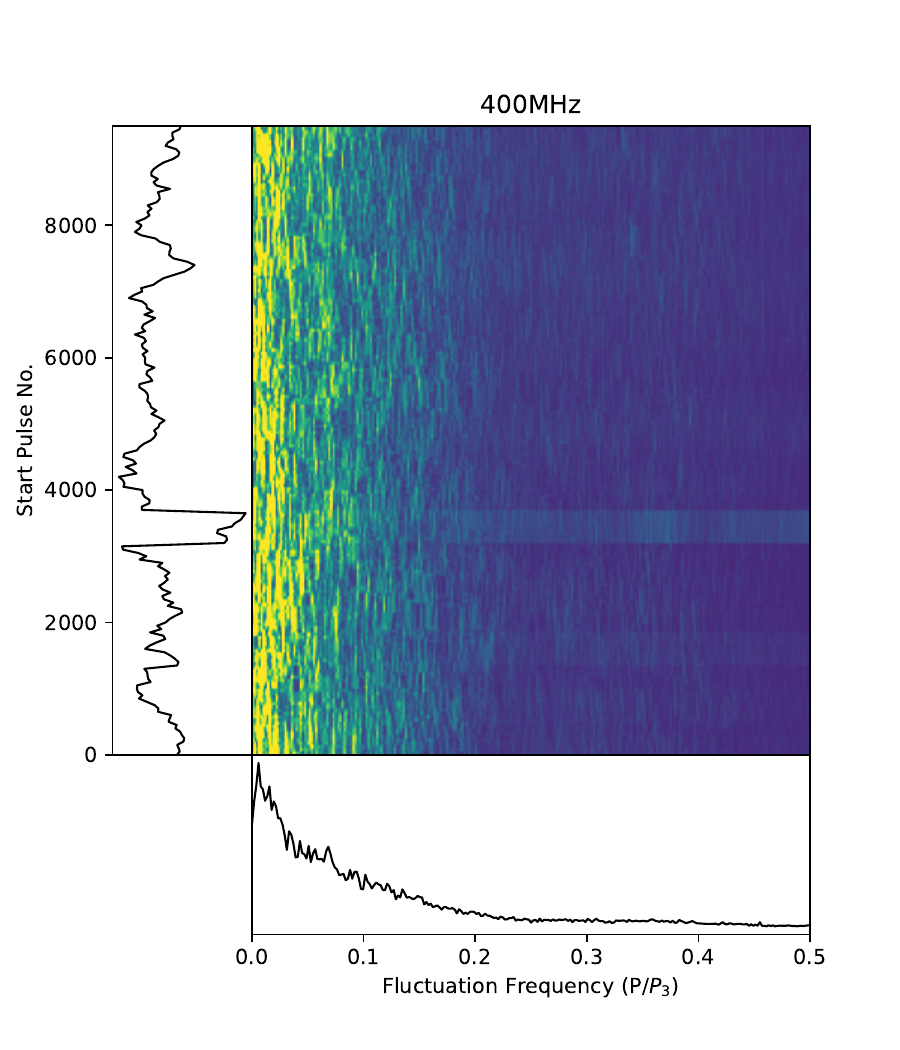}
    \includegraphics[width=0.32\linewidth]{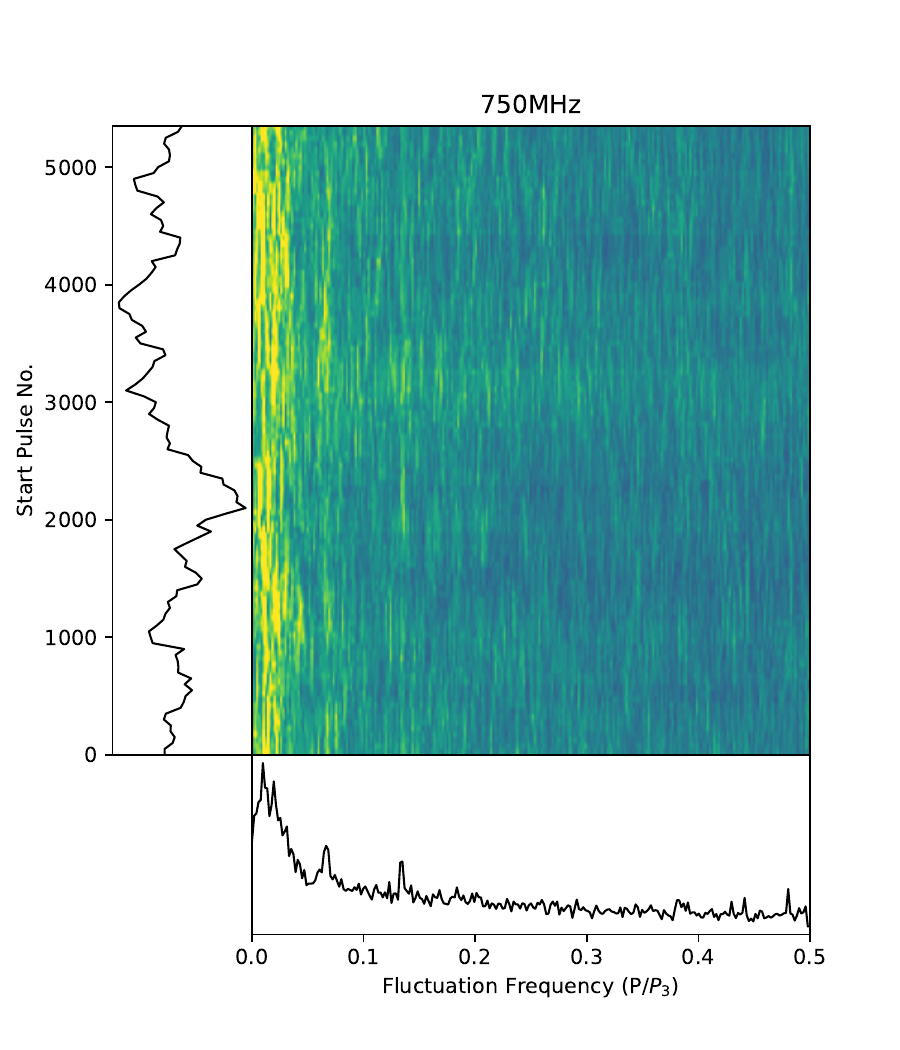}
    \includegraphics[width=0.32\linewidth]{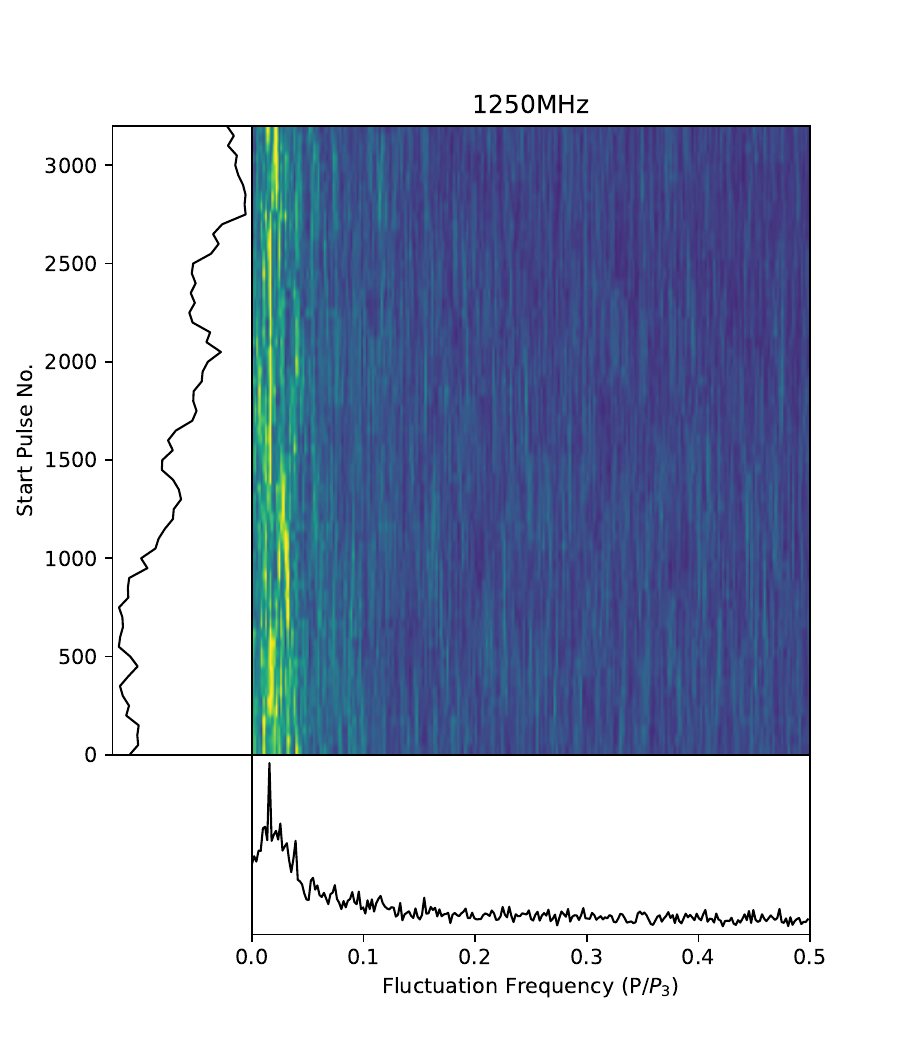}
    \caption{The time varying LRFSs of PSR J1948+3540 at $400\, \, \mathrm{MHz}$ (left), $750\, \, \mathrm{MHz}$ (middle) and $1250\, \mathrm{MHz}$ (right). The side panels of each column show the temporal variation of the LRFS. The bottom panels are the average LRFS, which are in units of $P/P_3$.}
    \label{fig:moveLRFS}
\end{figure*}

\begin{figure*}
    \centering
    \includegraphics[width=0.4\linewidth]{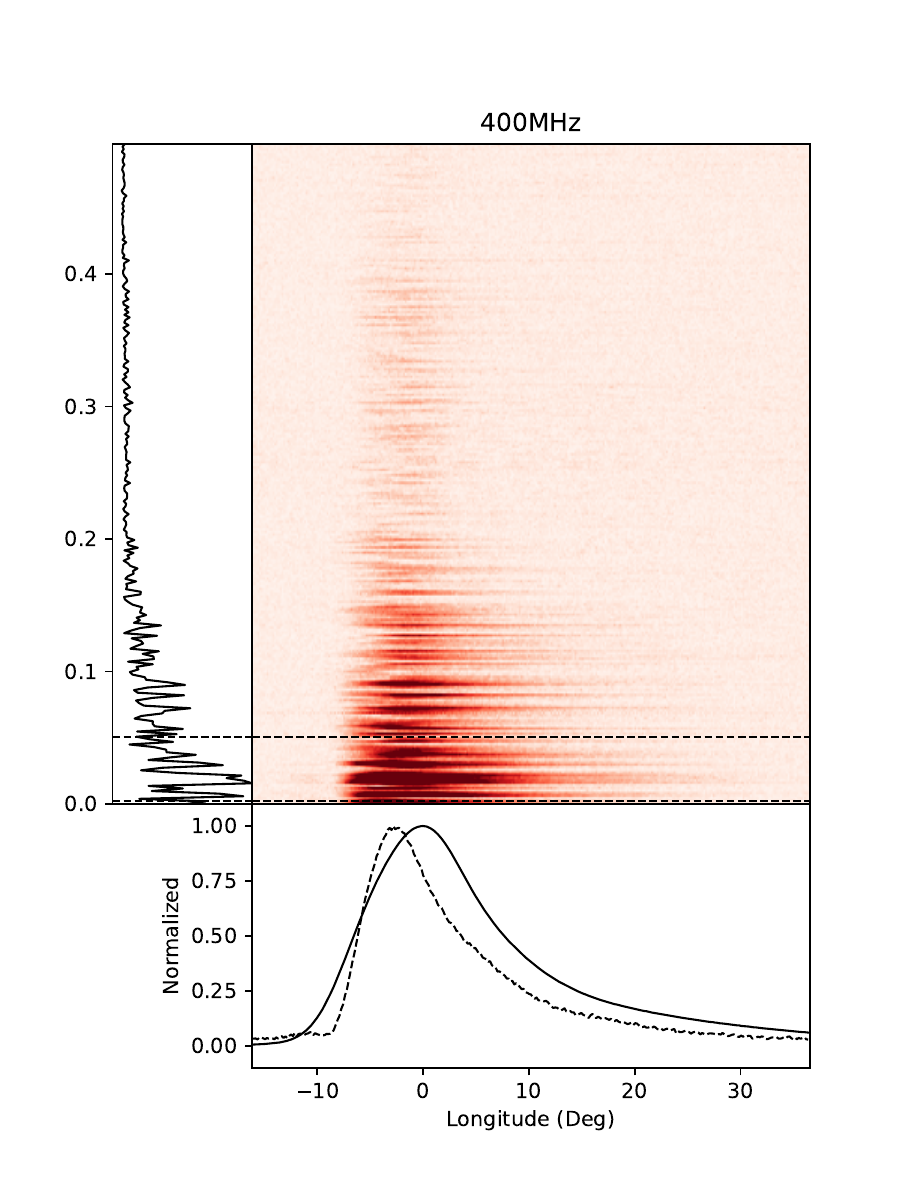}
    \includegraphics[width=0.4\linewidth]{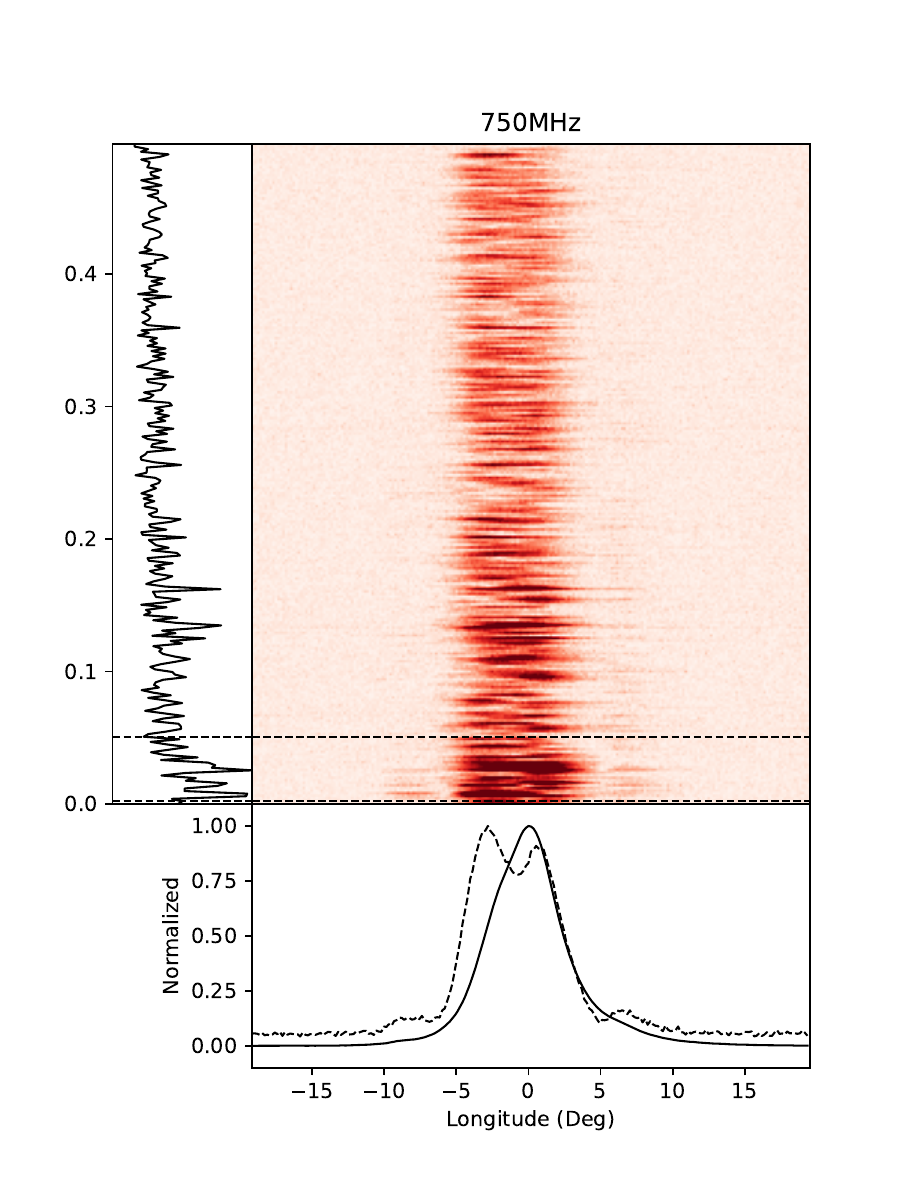}
    \includegraphics[width=0.4\linewidth]{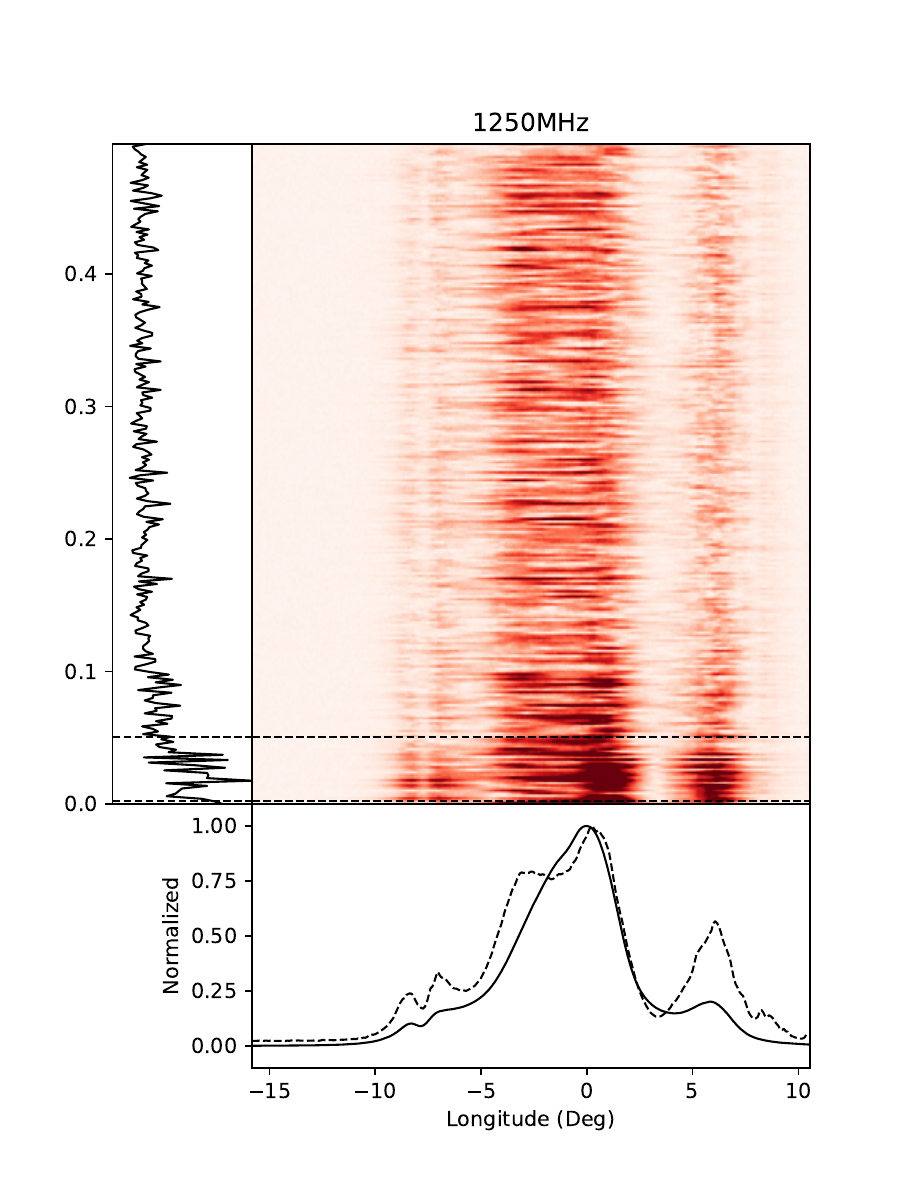}
    \includegraphics[width=0.4\linewidth]{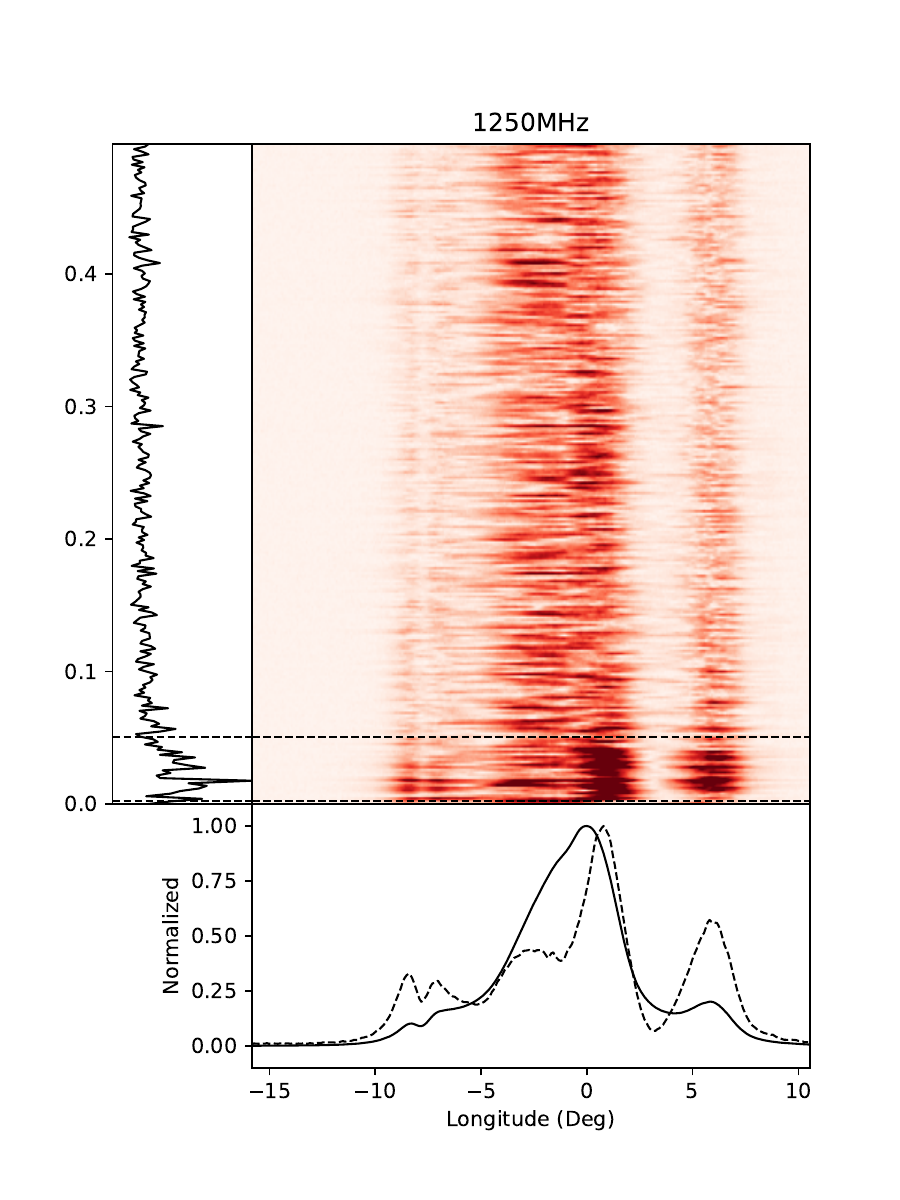}
    \caption{The fluctuation spectra analyses for different pulse blocks at center frequencies of $400\, \mathrm{MHz}$ (upper left),$750\, \mathrm{MHz}$  (upper right) and $1250\, \mathrm{MHz}$ (below). The upper rows are the results of the first $512$ pulses at $400\, \mathrm{MHz}$, $750\, \mathrm{MHz}$. The lower rows represents the pulse range from $600$ to $1112$ and the pulse range from $1500$ to $2012$ respectively. The left panels of each sub-figures show the LRFS and the bottom panels are the integrated pulse profiles. The vertical axis is the average LRFS, in units of $P/P_3$. In the bottom panels of each sub-plots, the black solid lines are the integrated pulse profiles, while the black dashed lines are the average power of LRFS at each longitude (bin) from $0.002 \, \rm cpp$ to $0.05 \,\rm cpp$ (the dashed lines in the upper panels).}
    \label{fig:LRFS}
\end{figure*}

To study the modulation in detail, we carry out fluctuation spectra analyses, which are shown in figures \ref{fig:moveLRFS} and \ref{fig:LRFS}. As reported in the \citet{mitra2017MNRAS.468.4601M}, the pulsar exhibits a time-varying amplitude modulation. We use the time varying Longitude Resolved Fluctuation Spectrum to check the stability of the modulation \citep{2016_Basu_Amp_flu}. This was done by calculating the Longitude Resolved Fluctuation Spectrum (LRFS for short) for each $512$-pulse block of the entire observation by shifting the starting point by $50$ periods. As we can see from the figure \ref{fig:moveLRFS} that the modulation of the pulsar exhibits broad low-frequency features with modulation frequencies range from $0.01\, \, \mathrm{cpp}$ to $0.05 \, \, \mathrm{cpp}$. At both $750\,\, \rm{MHz}$ and $1250\,\, \rm{MHz}$, it is evident that the modulation period of this pulsar varies with observation time. As reported in \citet{mitra2017MNRAS.468.4601M}, the intensity modulation is dominated by broad, low-frequency modulation with no clear periodicity for thousands of periods and then dominated by a relatively well-defined periodic modulation at $1.4\,\, \rm GHz$. We also observed similar phenomena in observations at $1250\,\, \rm MHz$. Therefore, in our subsequent analysis, we selected different pulse block for the data at $1250\,\, \rm MHz$. Since the modulation period changes over time, the values provided in the overall spectrum are only for comparison and reference. From the average LRFS, the peak frequency $f_{p}$  and the corresponding error $\delta{f_{p}}$ could be determined \citep{2016_Basu_Amp_flu}. They are respectively $0.0058\pm 0.0039 \, \rm cpp$, $0.019\pm0.014 \, \rm cpp$ and $0.01563\pm 0.0008 \, \rm cpp$ at three bands. In addition, at $750 \, \rm MHz$, there are also two significant modulation characteristics, peaking at $0.066\, \rm cpp$ and $0.15 \, \rm cpp$, respectively.    

LRFS analysis for different pulse block separately, which are plotted in Figure \ref{fig:LRFS}. The first $512$ pulses at $400\,\, \rm{MHz}$ and $750\,\, \rm{MHz}$ are selected. However, at $1250 \, \rm MHz$, pulses within the ranges of $600$ to $1112$ and $1500$ to $2012$ are selected for this analysis. Consistent with upper analysis, the pulsar primarily exhibits relatively broad low-frequency modulation. Therefore, some typical modulation periods at these three bands are listed for comparison and reference. At $1250 \, \rm MHz$, the peak frequency is $0.018\pm0.003 \, \, \rm cpp$ for pulses from $1500$ to $2500$, which implies a modulation period of $P_{3}\sim 56 P$. As for the first $1000$ pulses, they not only have a modulation with a period of $0.018 \, \rm cpp$, also exhibit another significant modulation component with a period around $0.023 \pm 0.05 \, \rm cpp $. We also performed the same analysis using GMRT data, yielding results that are in line with those obtained from FAST. However, there are differences in the modulation frequencies observed. Specifically, at $750\, \rm MHz$, the frequencies peak at $0.012 \pm 0.002\, \rm cpp$ and $0.021 \pm 0.003 \, \rm cpp$. While at $400\, \rm MHz$, the peaks are approximately $\sim 0.0058\, \rm cpp$ and  $\sim 0.0176\, \rm cpp$. The detailed results are list in Table \ref{tab:observation}. Taking statistical error into account, the modulation features with frequencies of $0.021 \pm 0.003 \, \rm cpp$ at $750\, \rm MHz$ and the $\sim 0.0176\, \rm cpp$ at $400\, \rm MHz$ are consistent with the two typical modulation frequencies observed at $1250\, \rm MHz$. In other words, these modulation features exist across all wavelength bands. Our results are largely consistent with those analyzed in other articles across different wavelength bands \citep{2006_Weltevrede,Weltevrede_2007,mitra2017MNRAS.468.4601M} . Indicating that the modulation frequency variation is time-dependent instead of frequency-dependent.

As shown in the Figure ~\ref{fig:LRFS}, at $750 \, \rm MHz$, the power of LRFS of $\rm{M_{I}}$ are higher than that of $\rm{M_{II}}$. However, at $1250\, \rm MHz$, the situation is reversed. This indicates that the intensity modulation is dominated by the first half of the middle component ($\rm{M_{I}}$) at low frequencies, while the latter half of this component ($\rm{M_{II}}$) becomes the primary contributor to the modulation at 1250~MHz. Additionally, at 400~MHz, the modulation is also dominated by the leading half of the pulse. We examined the spectrum published in the previous article and found that, in both the $1.4\, \rm GHz$ (from Arecibo) and $21\, \rm cm$ (from Westerbork Synthesis Radio Telescope) wavelength band data, the dominant modulation is in the latter half of the intermediate component \citep{mitra2017MNRAS.468.4601M,2006_Weltevrede}. However, at a wavelength of $92\, \rm cm$, the dominant modulation is in the in the first half of the pulse phase \citep{Weltevrede_2007}. Furthermore, since the modulation period/frequency is time-dependent, we are more inclined to believe that the dominate modulation components are frequency-dependent.

\subsection{The Phase Locking}
\begin{figure*}
    \centering
    \includegraphics[width=0.3\linewidth]{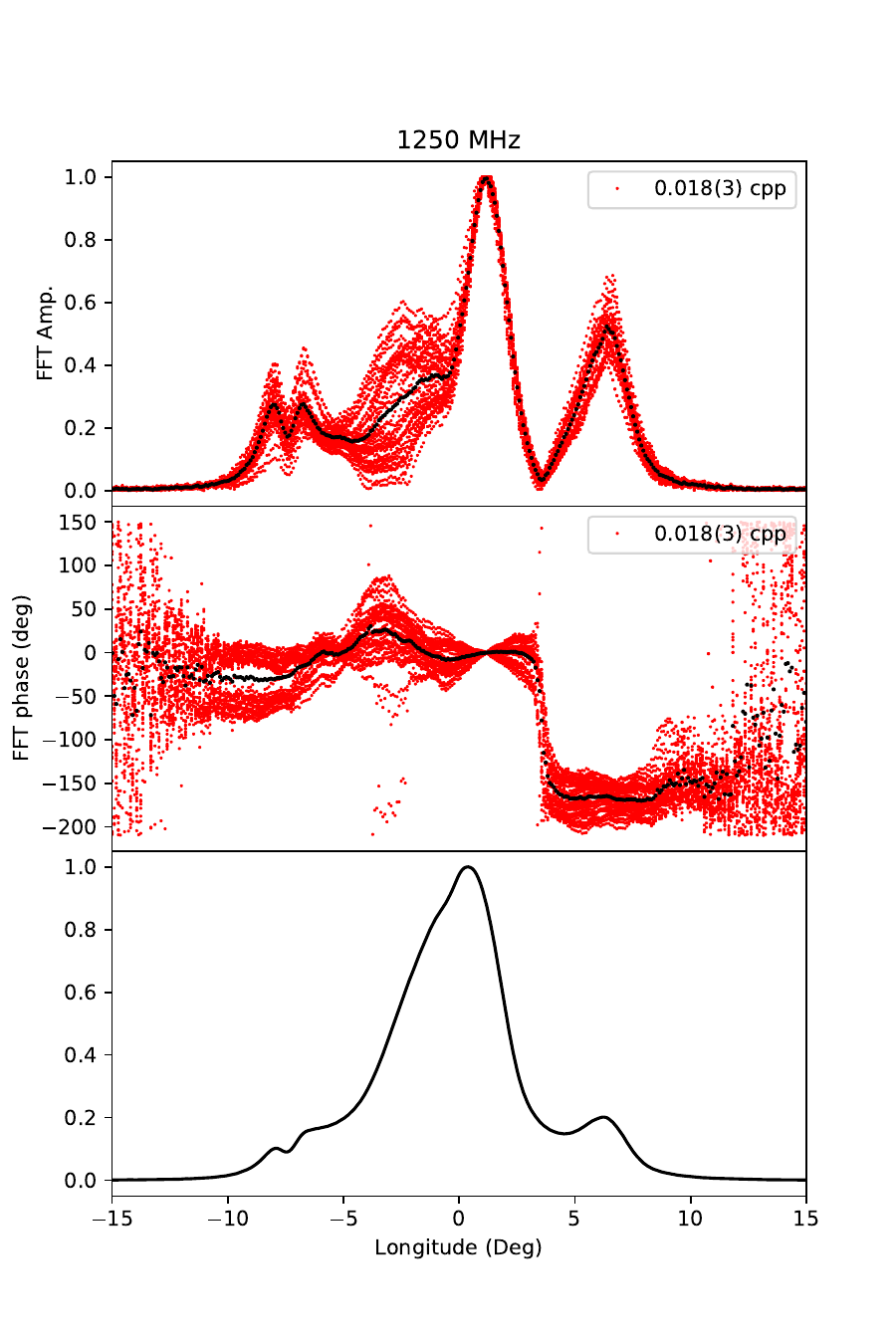}
    \includegraphics[width=0.3\linewidth]{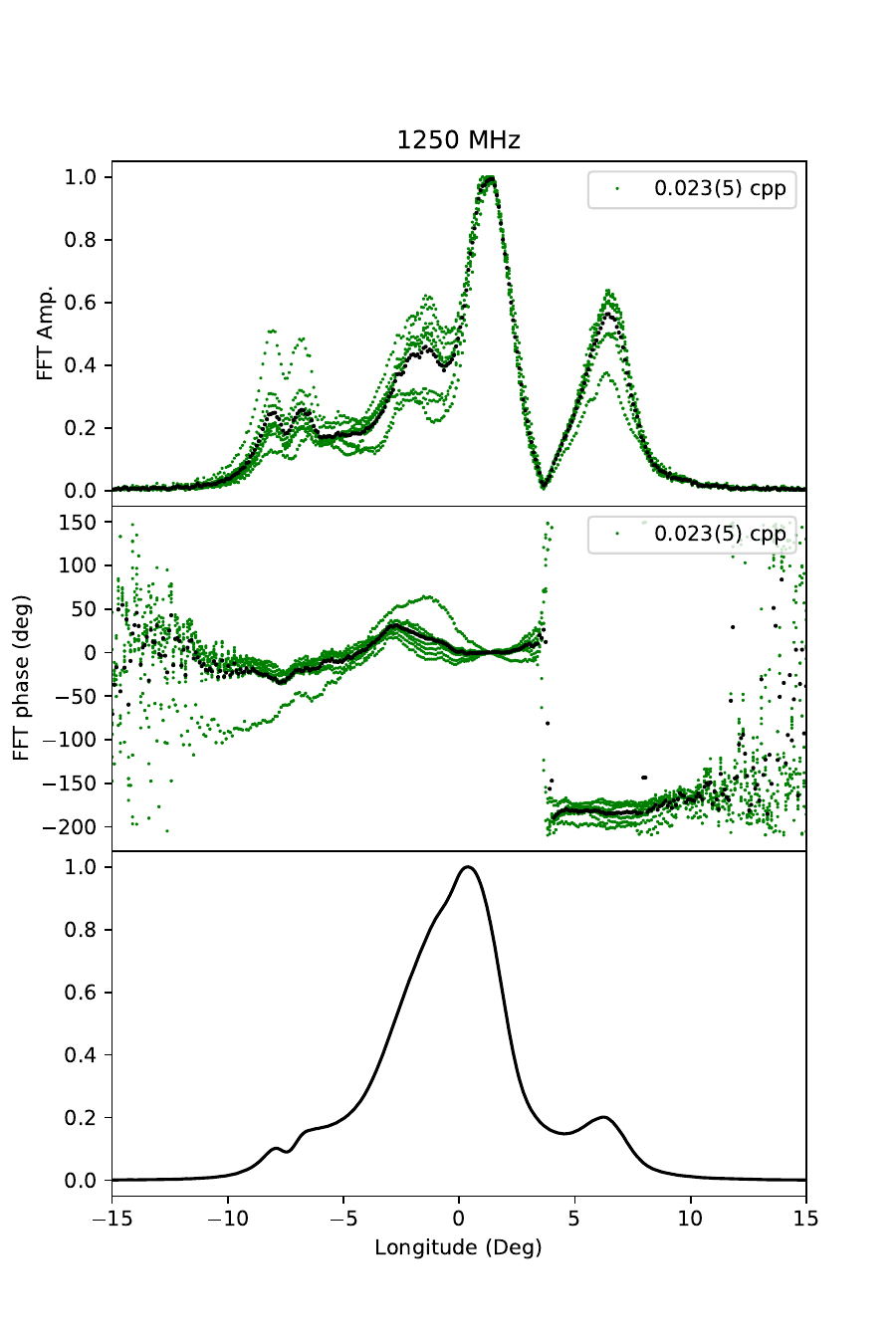}
    \includegraphics[width=0.3\linewidth]{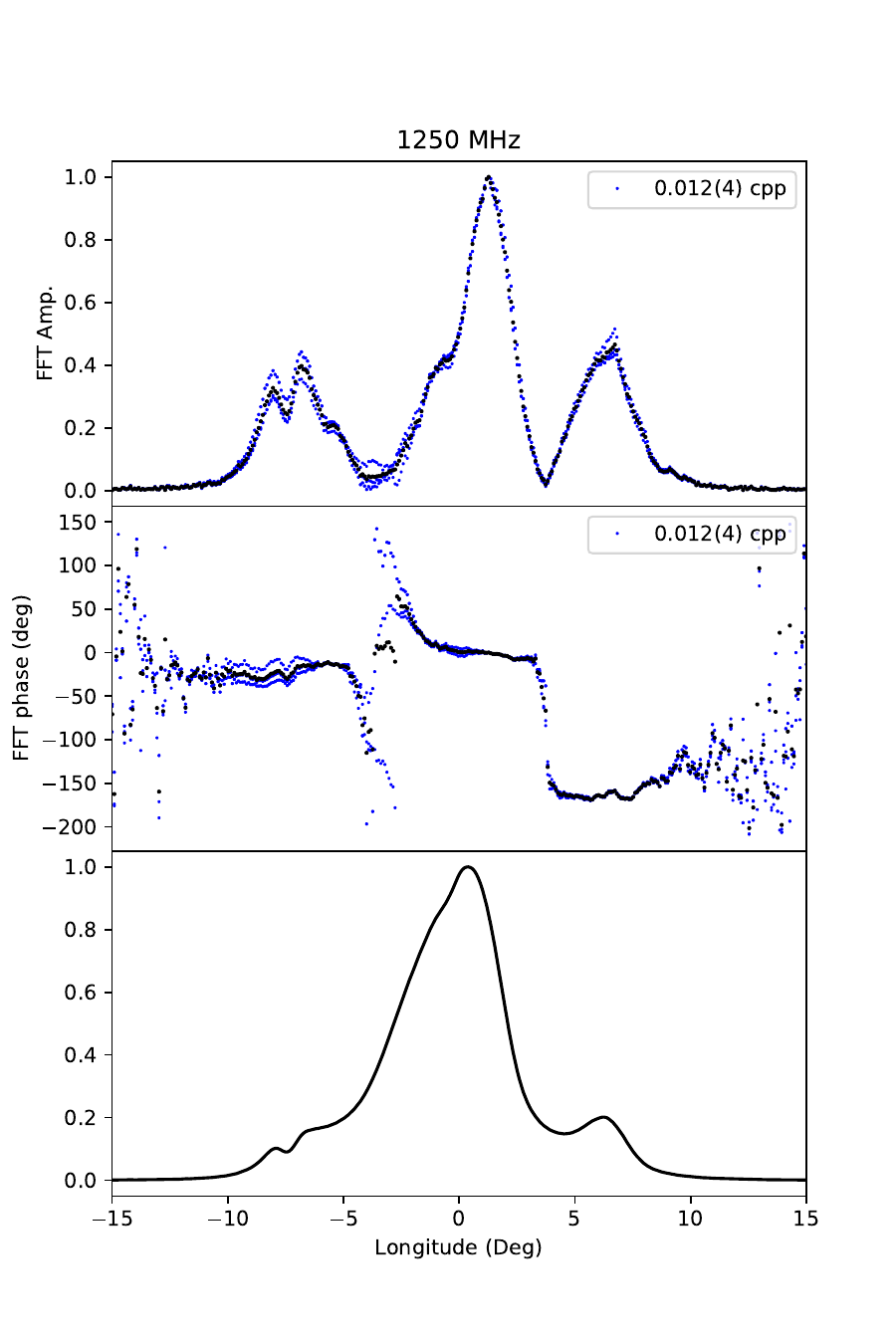}
    \caption{Phase variations of the LRFSs of the FAST data for different modulation frequencies of $0.018\, \rm cpp$ (left) and $0.023 \, \rm cpp$ (middle) and $0.012 \, \rm cpp$ (right). The panels from top to bottom show the peak amplitudes, the corresponding phase variations and the normalized integrated pulse profiles, respectively. The peak amplitudes and the phases for each LRFS are shown as red, green and blue points, while their average values are shown as black points.}
    \label{fig:fft-phase}
\end{figure*}

\begin{figure*}
    \centering
    \includegraphics[width=0.3\linewidth]{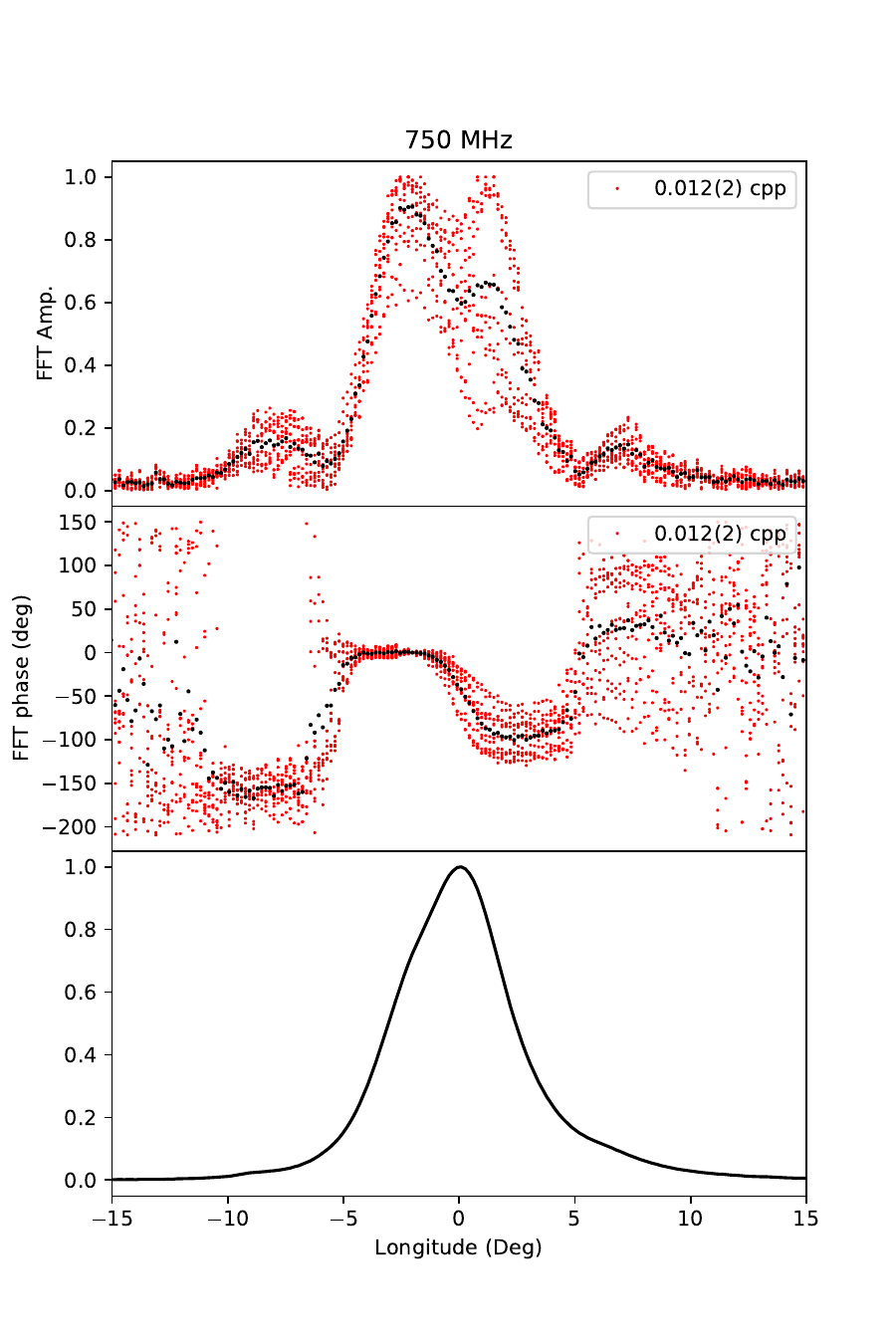}
    \includegraphics[width=0.3\linewidth]{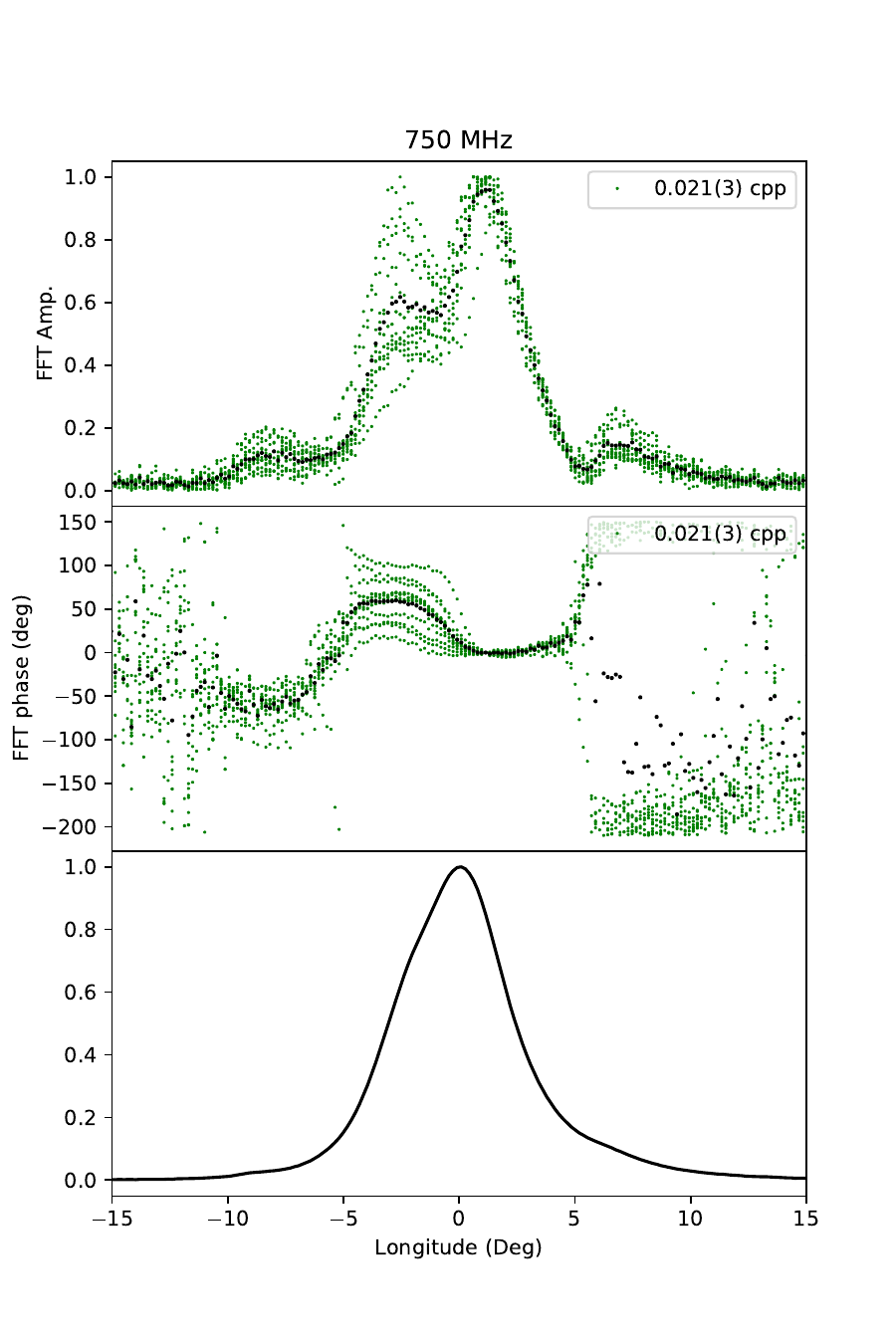}
    \includegraphics[width=0.3\linewidth]{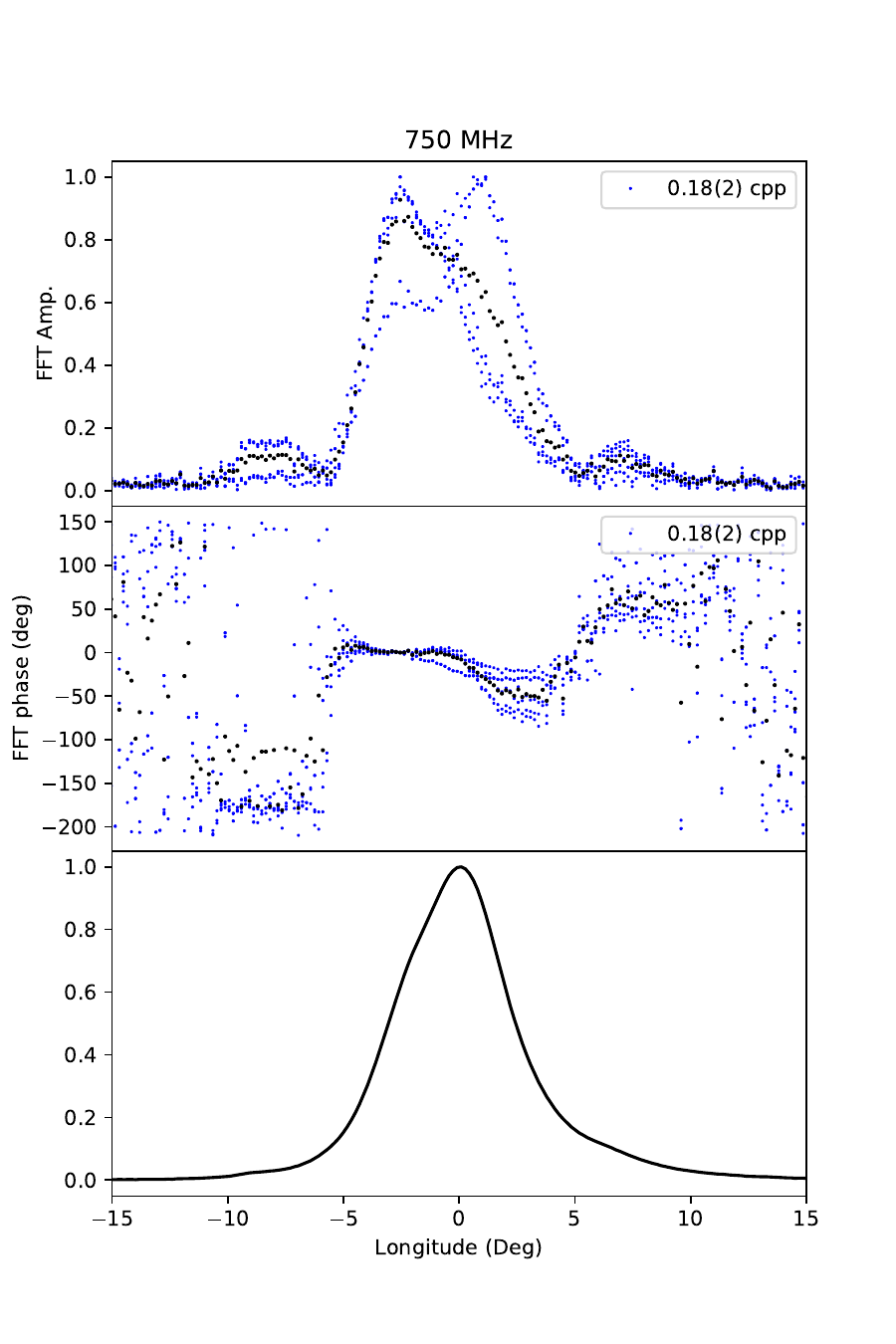}
    \caption{Phase Variations of the LRFSs of the GMRT data for different modulation frequencies, with left for $0.012\, \rm cpp$ , middle for $0.021 \, \rm cpp$ and right for $0.018\, \rm cpp$. The panels from top to bottom show the peak amplitudes, the corresponding phase variations and the normalized integrated pulse profiles. The peak amplitudes and the phases for each LRFS are shown as red, green and blue points, while their average values are shown as black points.}
    \label{fig:fft-phase2}
\end{figure*}

\begin{figure*}
    \centering
    \includegraphics[width=0.3\linewidth]{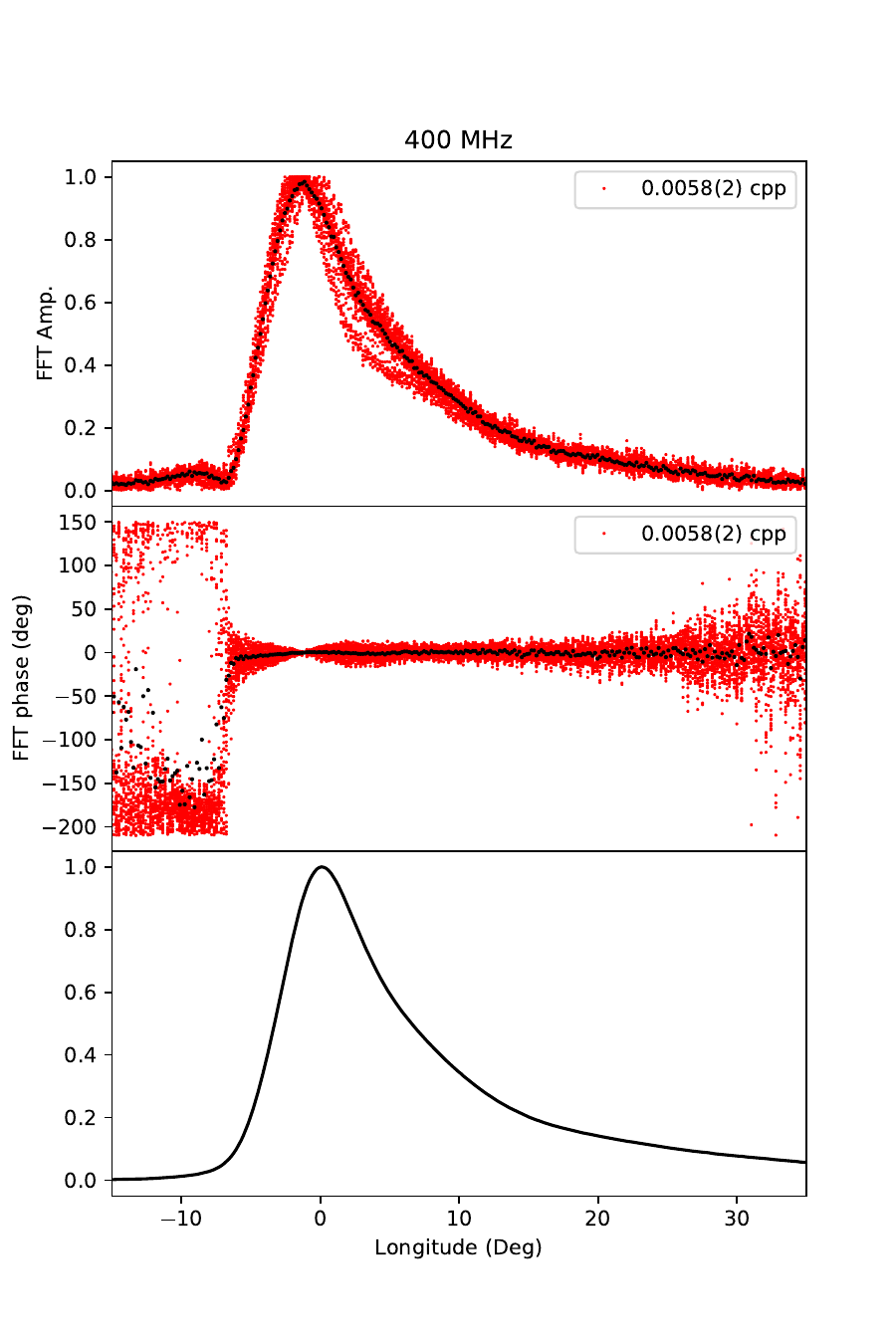}
    \includegraphics[width=0.3\linewidth]{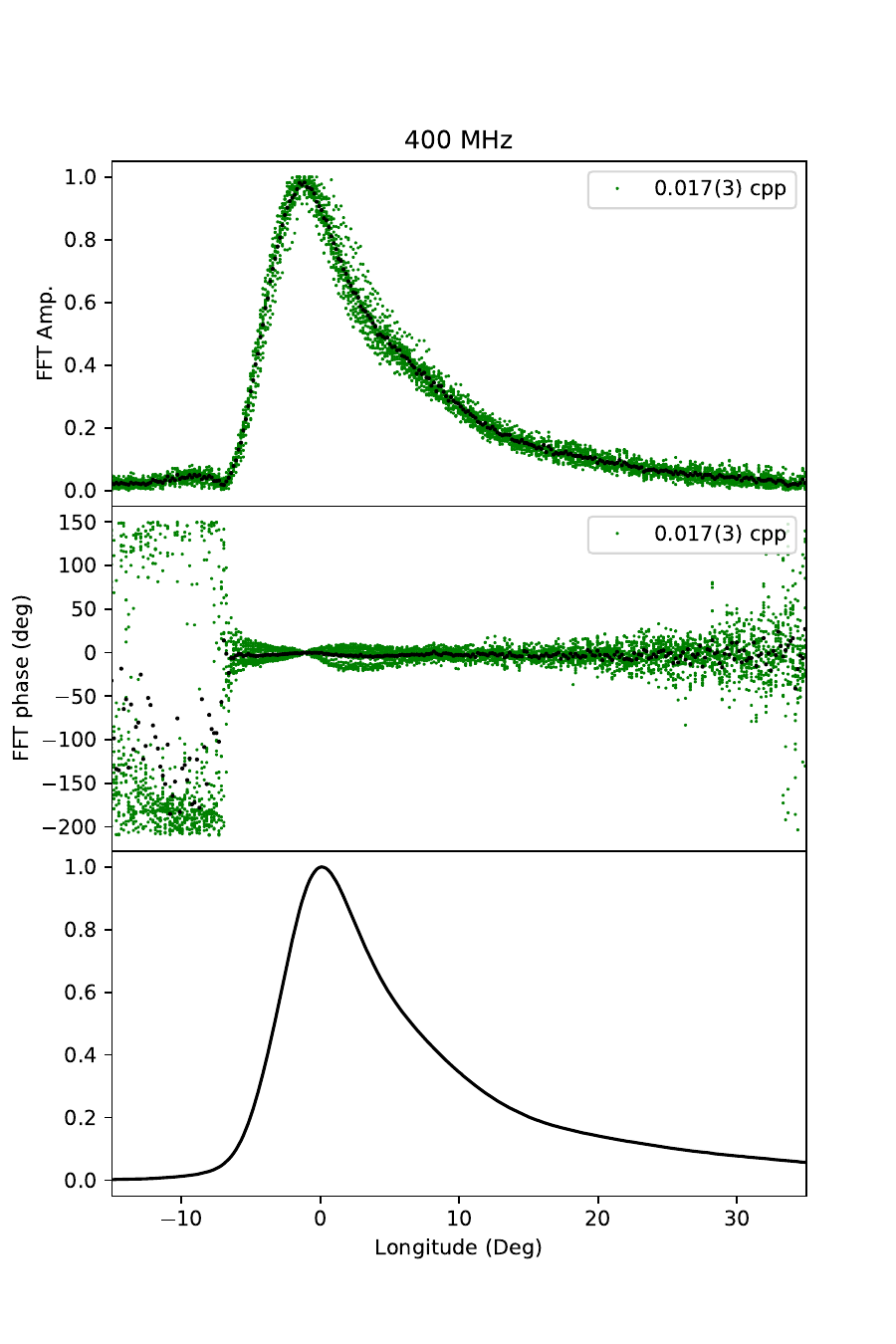}
    \includegraphics[width=0.3\linewidth]{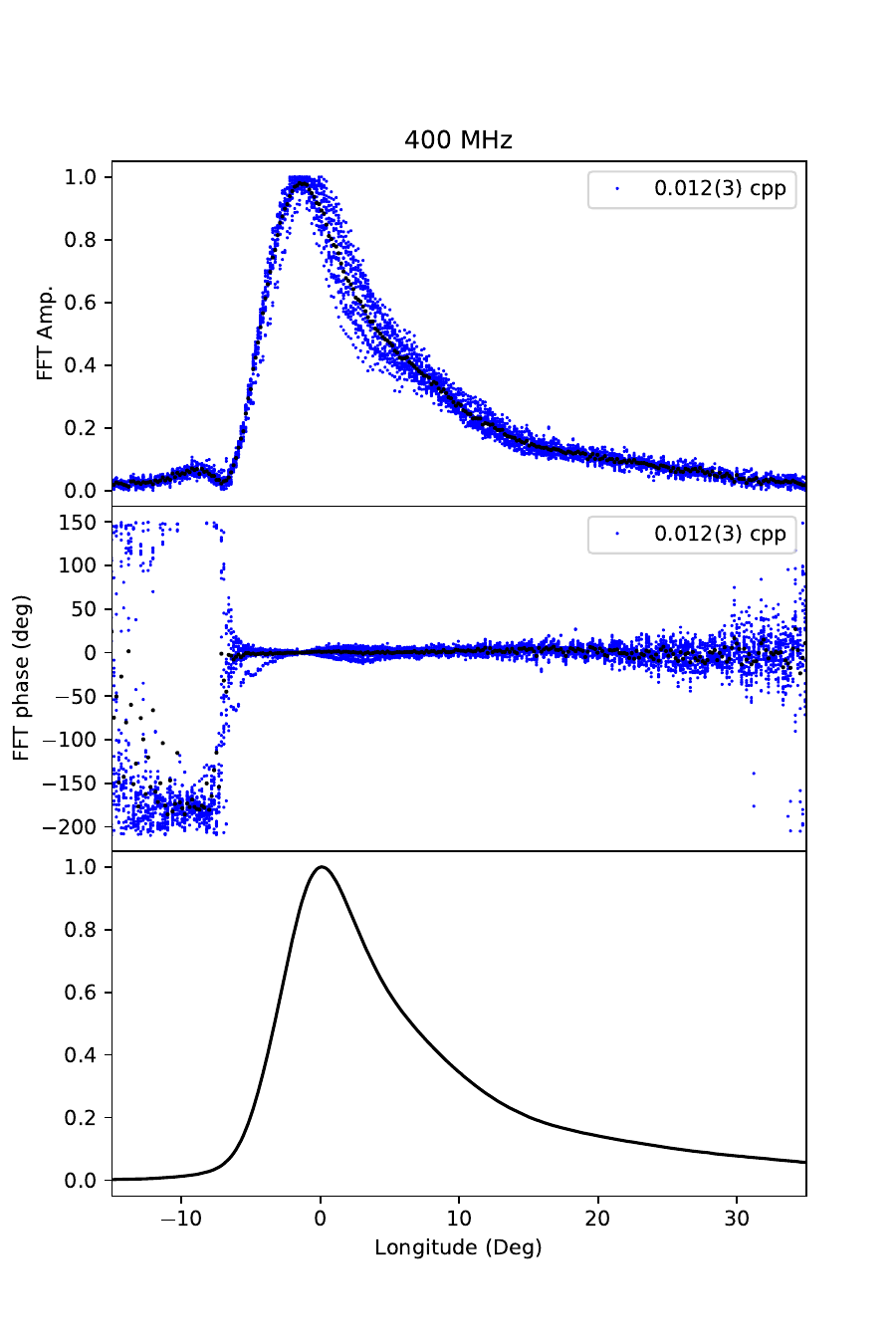}
    \caption{Phase Variations of the LRFSs of the GMRT data for different modulation frequencies, with left for $0.058\, \rm cpp$, middle for $0.017 \, \rm cpp$ and right for $0.012 \, \rm cpp$. The panels from top to bottom show the peak amplitudes, the corresponding phase variations and the normalized integrated pulse profiles. The peak amplitudes and the phases for each LRFS are shown as red, green and blue points, while their average values are shown as black points. }
    \label{fig:fft-phase-400M}
\end{figure*}

We conducted a further analysis of the phase variation to study the details of time-dependent intensity modulation and frequency-dependent dominate modulation components. Taking the data analysis at $1250\, \rm MHz$ as an example, which are plotted in Figure \ref{fig:fft-phase}. The analysis proceeded through the following specific steps: Firstly, we selected LRFSs with identical modulation frequencies/periods. Specifically, we chose two notable modulation frequencies: $0.018\, \rm cpp$ and $0.023\, \rm cpp$. For each of these modulation frequencies, we calculated the spectral amplitudes within each pulse bin and averaged them, resulting in the red and green points displayed in the top panels of the left and middle columns of Figure \ref{fig:fft-phase}. 
Secondly, to eliminate arbitrary phase differences among different blocks of LRFSs, we set the phase at the pulse longitude corresponding to the peak amplitude (which was $0.87^{\circ}$)—not the pulse longitude of the intensity peak ($0.0^{\circ}$) —to zero \citep{2018MNRAS.476.1345B}.  Subsequently, we estimated the phase differences, represented by the red and green points in the middle panels of the left and middle columns. The same analysis was done for the GMRT data at $400\, \rm MHz$ and $750\, \rm MHz$, which is shown in Figure \ref{fig:fft-phase-400M} and \ref{fig:fft-phase2}. At $750 \, \rm MHz$, the main modulation frequencies are $0.012\, \rm cpp$ and $0.021\, \rm cpp$, and the pulse longitudes corresponding to the peak amplitudes are $-3^{\circ}$, and $0.87^{\circ}$, respectively. In fact, even though the modulation centered at $0.021\, \rm cpp$ comes later, we can still observe a strong modulation intensity in the first half compared with those at 1250~MHz. For comparative analysis, we also screened the modulation frequencies of $0.012\, \rm cpp$ at 1250~MHz, and the modulation frequencies of $0.018\, \rm cpp$ at 750~MHz. The results remain consistent with previous: low-frequency modulation is predominantly governed by $\rm{M_{I}}$, whereas $\rm{M_{II}}$ emerges as the dominant modulation parts at 1250~MHz. Besides, the variation of the dominant modulation component is frequency-dependent. As supplementary data, the 400~MHz measurements reveal a dominant modulation frequency of $0.0058~\rm cpp$.  We also selected $0.018~\rm cpp$ and $0.012~\rm cpp$ for comparative analysis. The modulation centers are all at $-3.5^{\circ}$, indicating that the modulation is  dominated by the leading half of the pulse.

As shown in the figures (\ref{fig:fft-phase} and  \ref{fig:fft-phase2}), the components with same modulation frequencies are phase-locked. We did not conduct phase-lock analysis on the 400~MHz data because they were severely affected by scattering.
There are time delay between their intensity modulation. Taking the FAST data with modulation frequency of $0.018 \, \rm cpp$ as an example, the phase difference between the leading component and the middle component can indeed be categorized into two distinct groups: one group is close to $0^{\circ}$, and the other is close to $-60^{\circ}$. The phase difference between the trailing component and the main component is about $-160^{\circ}$. This indicates that the intensity fluctuation of the leading component changes in synchronization with the intermediate component or lags behind it by approximately $5$ periods. However, the trailing component lags behind the main component about $24$ periods. The detailed analysis results for the three bands are presented in Table \ref{tab:observation}. The radiation components, despite sharing the same modulation period, exhibit non-synchronous intensity variations. 
The time delays make it difficult to distinguish emission modes.

\subsection{The Spectrum}
The pulsar exhibits significant frequency-dependent features: the intensities of the leading and trailing components increase as the observation frequency increases, while that of the middle component decreases; Meanwhile, the component that dominates the periodic modulation also changes with frequency. We conducted a Phase-resolved Spectrum analysis in this subsection. 

\begin{figure}
    \centering
    \includegraphics[width=70mm]{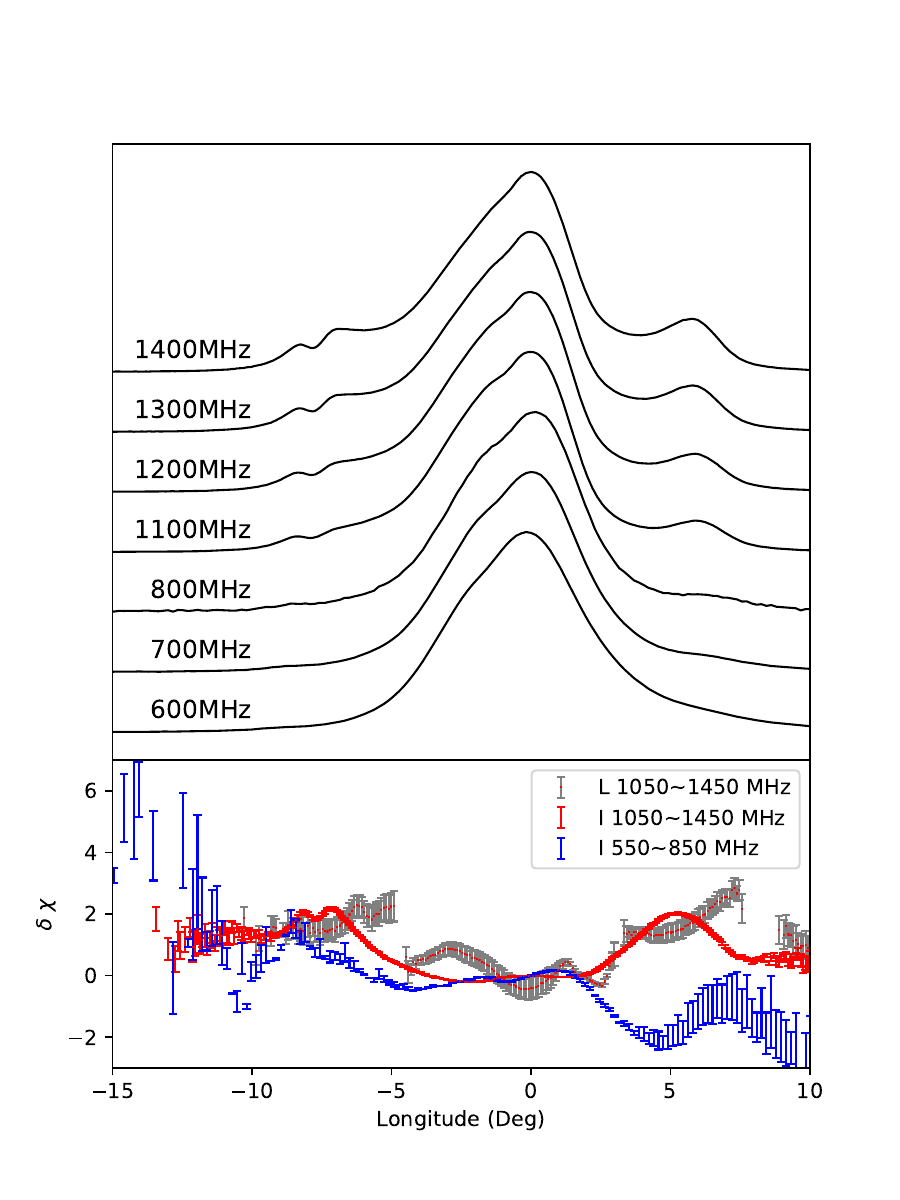}
    \caption{The normalized pulse profile at different frequencies (upper panel) and longitude-resolved spectral indexes (lower panel). The red points with error bars of the same color are the longitude-resolved spectral indexes calculated for the total intensity at frequencies ranges from $1050\, \rm MHz$ to $1450\, \rm MHz$. The blue points are for frequencies from $550\, \rm MHz$ to $850\, \rm MHz$. The red points with gray error bars are the longitude-resolved spectral indexes calculated for the linear polarization at frequencies ranges from $1050\, \rm MHz$ to $1450\, \rm MHz$.}
    \label{fig:delx}
\end{figure}

The FAST data and and GMRT data at  band $4$ were respectively divided to $8$ channels and $4$ channels to calculate the phase-resolved spectral index \citep{2007_Chen,2024_Cai}. Since the data has not undergone flux calibration, we perform the calculations separately for the data from the two observations. Assuming that the intensity evolution of pulsars follows a power-law spectrum as a function of the observing frequency $I=K \nu^{\chi}$. We selected the pulse phase with the highest radiation intensity as the reference phase, and divided the radiation intensity at all pulse phases by the intensity at this reference phase, and got the corresponding intensity ratio $\eta_i=I_i/I_{0}=K_i/K_{0} \nu^{\chi_{0}-\chi_i}=K_i/K_{0} \nu^{\delta \chi}$. Subsequently, the phase-resolved spectra could be calculated by fitting the data with equation $\log \eta_i=\delta \chi_i \log \nu +C_i$, where $C_i = log(K_i/K_0)$. The normalized pulse profile at different frequencies and longitude-resolved spectral indexes are plotted in Figure \ref{fig:delx}. The red and blue points are respectively results of the FAST data (from $1050\, \rm MHz$ to $1450\, \rm MHz$) and the GMRT data (from $550\, \rm MHz$ to $850\, \rm MHz$). Due to scattering effects, the spectral index of trailing component is distorted (this is also the reason why we discarded the data from band$3$ in this analysis.).
Consequently, the data results obtained from GMRT can only serve as supplementary evidence to support the results made by FAST.
Here, it can be seen more clearly that as the observation frequency increases, the relative intensities of $Com_L$ and $Com_T$ are also increasing.
The spectral indexes of the leading and trailing components are relatively higher than those of the middle component. This indicating that the spectrum of the leading and trailing components are much flatter that that of the middle component. This is the reason why the radiation intensity of these two components is relatively enhanced at high frequencies. The spectral index values of the middle components are basically the same, indicating that their intensities change consistently with the observing frequency. The variation in the dominant modulation component with observation frequency is not caused by changes in intensity. 

\begin{figure}
    \centering
    \includegraphics[width=1.0\linewidth]{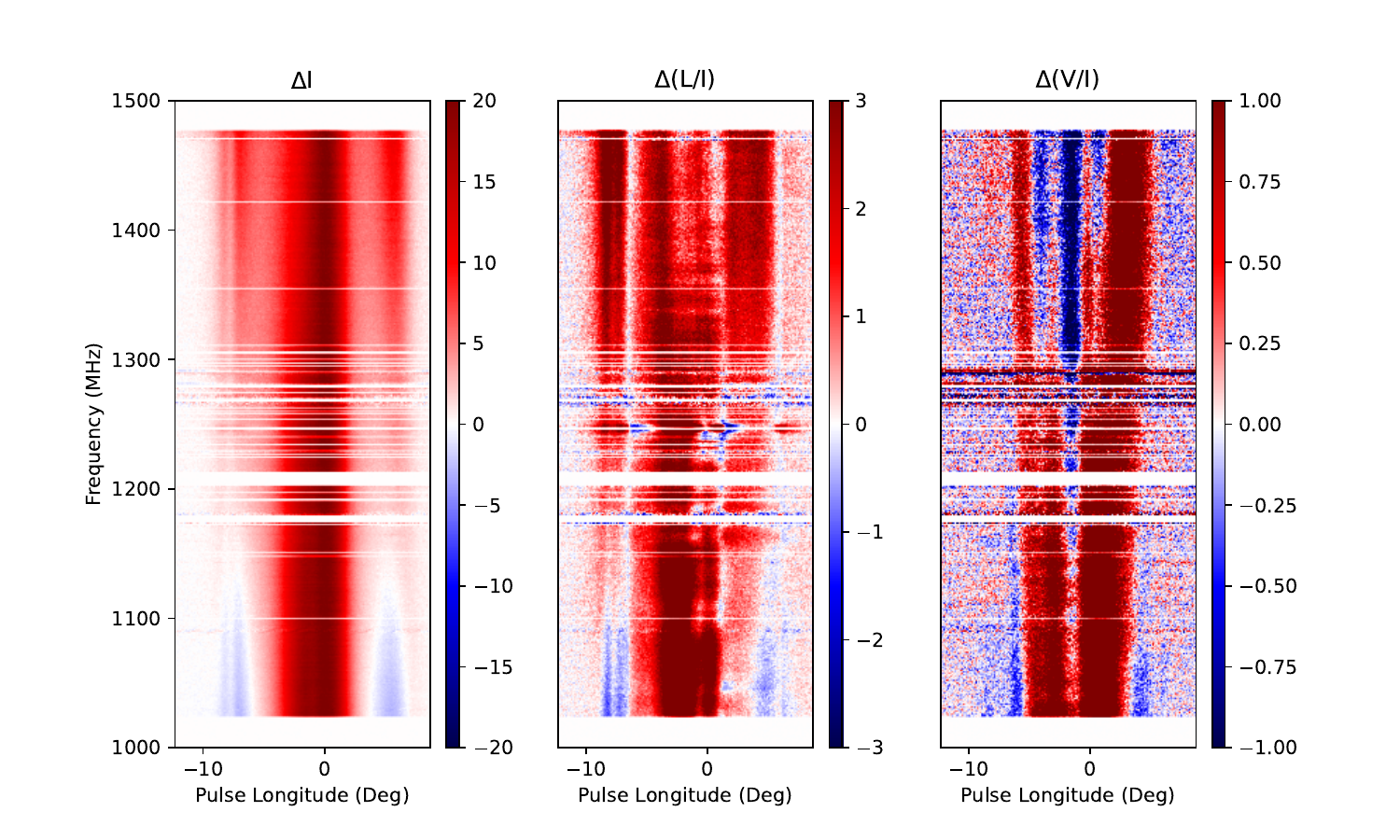}
    \caption{The smoothed difference evolution of the total intensity, linear polarization intensity and circular with frequency from $1050\, \rm MHz$ to $1450\, \rm MHz$. The colors corresponding to different intensities are displayed in the color-bar on the right. The white line in the middle part of each panels is the interference band marked and eliminated.}
    \label{fig:frequency-1250}
\end{figure}

The variation of the polarization profiles with the observation frequency of the FAST data is shown in Figure \ref{fig:frequency-1250}. As we can see that, the total intensity and the linear polarization intensity of $\rm{Com_I}$ and $\rm{com_{T}}$ increases with observational frequency. Meanwhile, those of $\rm{M_{I}}$ and $\rm{M_{II}}$ decreases. The circular polarization intensities of $\rm{Com_I}$ and $\rm{com_{T}}$ are relatively weak, and no significant changes in circular polarization intensity were detected. However, the circular polarization of $\rm{M_{I}}$ is enhanced, while that of $\rm{M_{II}}$ reduced. Besides that, the peak of the radiation intensity is located in the latter half of the middle component ($\rm{M_{II}}$), whereas the linear polarization is dominant in the former half ($\rm{M_{I}}$), which is consistent with the results of \citet{mitra2017MNRAS.468.4601M}. We conducted a spectral analysis for the linear polarization using the FAST data, the results are presented as red points accompanied by gray error-bars in Figure \ref{fig:delx}. The linear polarization spectrum of $\rm{M_{I}}$ is relatively flat compared to $\rm{M_{II}}$). Additionally, the linear polarization degree of this pulsar is very low, it is not the variation in linear polarization that leads to the transition of domain modulation components with frequency. 

\begin{figure}
    \centering
    \includegraphics[width=0.8\linewidth]{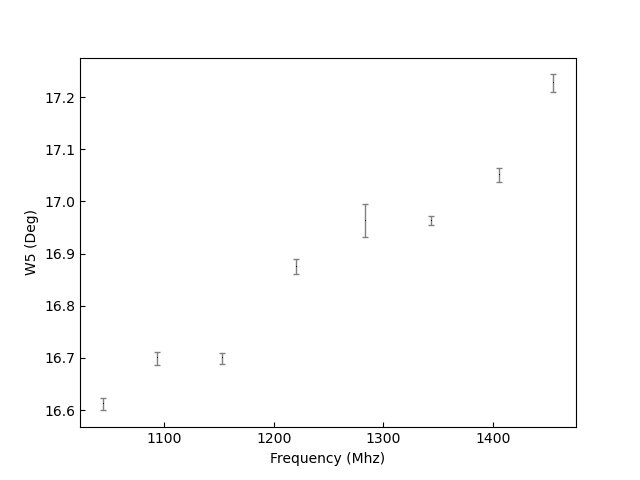}
    \includegraphics[width=0.8\linewidth]{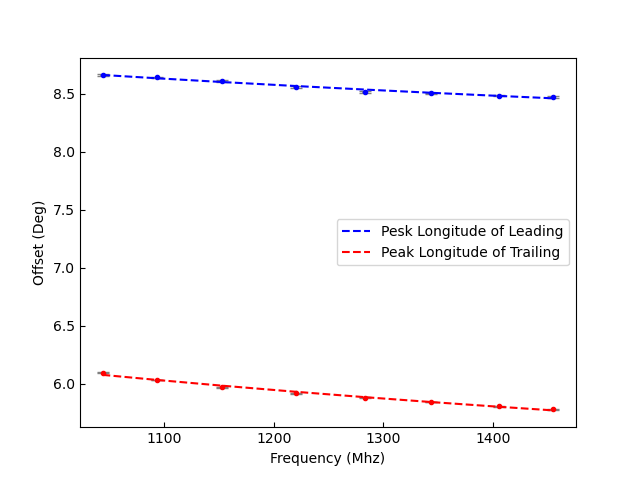}
    \caption{The pulse width and pulse offset evolution with frequency. The red and blue points in the lower panel are respectively the pulse offsets of the leading and trailing components, and the dashed lines are the best-fit results.}
    \label{fig:pulsewidth}
\end{figure}

The pulse width, as well as the pulse offset of the leading component and trailing component, are shown in Figure \ref{fig:pulsewidth}. Because the radiation of the precursor and trailing components is relatively weak, the pulse width at the $5\%$ height of the profile is calculated. To accurately locate the different peak longitudes and their errors, we fit the profiles of the pulsar at different frequency bands using multi-Gaussian functions. As we can see that, though the pulse width increases as frequency, the pulse offsets of the leading component and the trailing component decrease with frequency. The frequency dependence of the offset is consistent with the radius-to-frequency mapping. Based on previous statistical results, frequency dependence of the offset follows a power-law. We derived the power-law index for the offsets of the the leading component and the trailing component, which are $0.071(4)$ and $0.156(6)$ respectively. As for the broadening of the profile with frequency, it is apparently caused by an increase in relative intensity.

\subsection{The Polarization}
\begin{figure}
    \centering
    \includegraphics[width=0.9\linewidth]{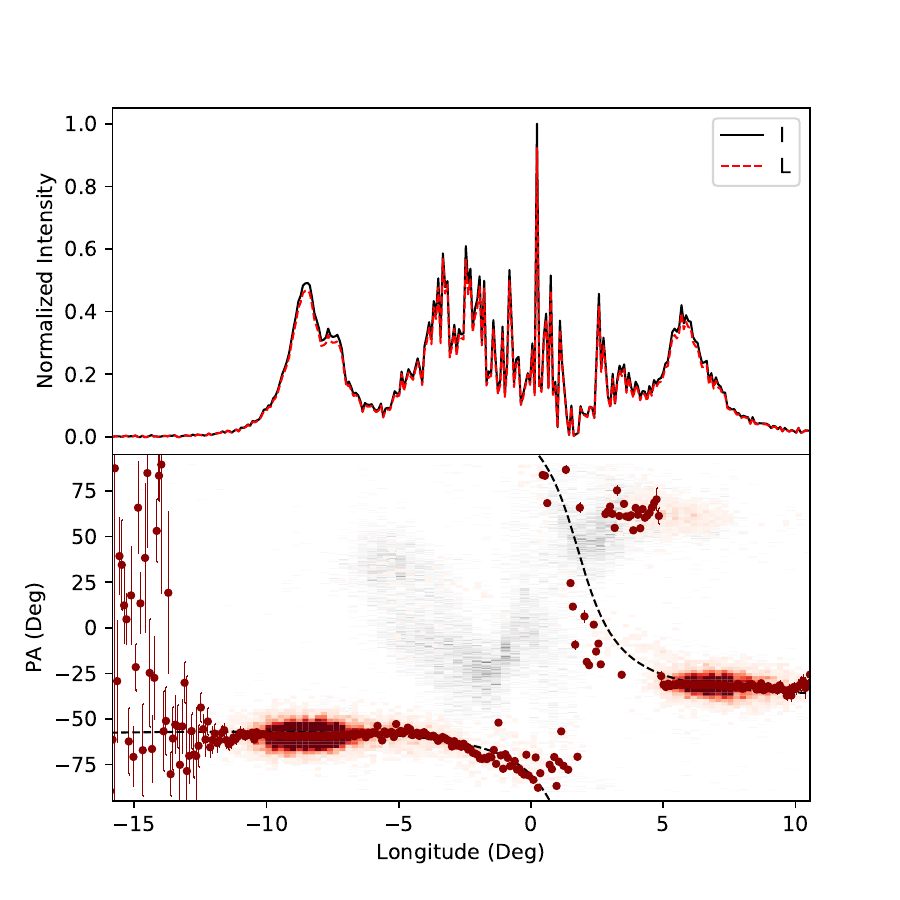}
    \caption{ The integrated pulse profile (top panel) and PA distribution (bottom panel). In the top panel, the black and red dashed lines represent total intensity and linear polarization intensity (L) of highly polarized single pulse time samples ($L/I>=90\%$). In the bottom panel, the grey shaded region is the PA distribution of total single-pulses, while the red region shows the PA distribution of highly polarized single pulse time samples and their average PAs are shown as darkred points. The black dashed line is the the RVM fits to the average PAs. The geometric parameters used for RVM fits are $\alpha=142.48^{\circ}$, $\zeta=141.77^{\circ}$, $\psi_{0}=43.21^{\circ}$ and $\phi_{0}=1.77^{\circ}$.}
    \label{fig:PA-distribution}
\end{figure}

As demonstrated in Figure \ref{fig:prof}, the position angle (PA) profile of this pulsar exhibits a complex morphology characterized by multi-harmonic components and significant deviations from the canonical Rotating Vector Model (RVM) predictions. Besides, we found that the linear polarization degree of the conal component is much higher than that of the core component. Recent works have found that
the PA distribution of highly polarized time samples in the single pulses clearly follows the RVM \citep{2023_Basu,2023_mitra_1645,2024_Johnston}. Following the method in \citet{2023_mitra_1645}, we also attempted to select highly polarized time samples in the single pulses and conduct RVM fitting, the results are shown in Figure \ref{fig:PA-distribution}. The highly polarized time samples mainly from the leading and trailing components. We performed Markov Chain Monte Carlo fitting \citet{2019MNRAS.490.4565J} using the python package EMCEE \citet{Foreman-Mackey_2013} to the PA curve and the best fiting results are $\alpha={142.48^{\circ}}^{+16.06}_{-16.11}$, $\zeta={141.77^{\circ}}^{+16.36}_{-16.37}$, $\psi_{0}={43.21^{\circ}}^{+0.41}_{-0.41}$ and $\phi_{0}={1.77^{\circ}}^{+0.11}_{-0.10}$.

The PA distribution of total single-pulses exhibits a V-shaped, and the lowest point in phase with the lowest point of the circular polarization of the middle component. \citet{mitra2017MNRAS.468.4601M} speculated that this phenomenon is caused by aberration/retardation because of the different pulse longitudes between the radiation center and the linearly polarized radiation center. Despite the significant overlap in pulse longitude and the general similarity in spectrum between $\rm{M_{I}}$ and $\rm{M_{II}}$, they display distinct modulation characteristics as the observing frequency changes. The complex radiation and polarization properties of the middle component remain poorly understood.

\section{DISCUSSION AND CONCLUSIONS} \label{section:4}
The radiation of this pulsar exhibits extremely complex and fascinating characteristics, summarized as follows: $1)$ it exhibits broad low-frequency modulation features with the modulation frequency being time-dependent; $2)$ the radiation components with same modulation frequencies are phase-locked; $3)$ the dominant modulation component shifts from the first half to the second half of the middle component as frequency increases, and this is not due to differences in the spectrum; $4)$ the spectra of the leading and trailing components are flatter than those of the middle component leading to a relative enhancement of high-frequency radiation, which indicate radiation origin distinct from the middle component; $5)$ the PA curve exhibits a complex V-shape/hook shape. These phenomena make it extremely difficult to simulate the magnetospheric geometry. 

\subsection{The Time-dependent Modulation}
Time-dependent modulations are a common phenomenon in pulsar observations, especially in drifting pulsars. The modulation in PSR J$1948+3540$ is taken as amplitude modulation instead drifting, because no significant drifting characteristics have been detected.

Observations also reveal that in mode-changing pulsars, different modes display distinct pulse profiles and modulation periods. For example, PSR B$0823+26$ switches between its Bright-mode and Quiet-mode according to the appearance and disappearance of the precursor component and inter-pulse \citep{Basu_2019}. It also exhibit different modulation period and nulling fraction in its B-mode and Q-mode \citep{Basu_2019}. Similar phenomena have also been observed in PSR B$1822-09$, which also show different modulation periods during its  B-mode and Q-mode \citep{Latham_2012,2019_Yan_1822}. It is believed that the periodic amplitude modulations and nulling observed are the results of a triggering mechanism, which operates within the pulsar's magnetosphere to periodically alter the pair production process \citep{Basu_2019}. However, the triggering mechanism remains unknown. 

\subsection{The Frequency-dependent Modulation Features}
However, the frequency-dependent modulation is only reported in the observations of PSR B$0031-07$ \citep{Huguenin_1970,McSweeney_2017}. The pulsar is well known to exhibit three different drifting sub-pulse modes at low frequencies, and only one mode is visible at high frequencies \citep{Smits_2005,Smits_2007}. A geometrical model was proposed based on the observation in which two nested concentric rings emit modes A and B at a given frequency, with mode B located inside mode A. As the observing frequency increases, at heights closer to the star, the line of sight only intersects the outermost radiation ring, resulting in the visibility of only mode A \citep{Smits_2005,Smits_2007}. 

In addition to not being a subpulse drifting pulsar, PSR J$1948+3540$ also exhibits differences in how its modulation features vary with observing frequency compared to PSR B$0031-07$. As the observing frequency changes, the dominance of the modulation components changes, but the radiation component does not disappear. Therefore, it may not be possible to explain this phenomenon simply by differences in the height of the emission origin. 

\subsection{The Shifted-pulses}
The radiation of PSR J$1948+3540$ also show both similarities and differences compared to the pulse shiftting pulsars J$0922+0638$ (B$0919+06$), J$1901+0716$ ($B1859+07$) and J$0614+2229$ (B$0611+22$). These pulsars are well-known for their sudden shifts in pulse emission toward earlier longitudes over several pulses, followed by a return to their normal emission phases \citep{Rankin2006,rajwade2021MNRAS.506.5836R,sun2022ApJ...934...57S,2024_Cai}. Just like the shifted pulses, the leading and trail components of PSR J$1948+3540$ alternate in appearance, with their modulation periods remaining consistent with those of the main radiation component. Besides, the intensity of the leading and trail components increases as the observation frequency increases, while that of the middle component decreases. Meanwhile, our observations have also revealed that the evolution of the intensity and pulse width of these two components is similar to that of these three pulsars. Specifically, as the observing frequency increases, the relative radiation intensity enhances, and although the pulse profile broadens, the offset of the shifted pulses decreases. \citet{rajwade2021MNRAS.506.5836R} had proposed a competitive model that the shifted pulse may originate from a higher altitude, and with the shrinking and expanding the magnetosphere, the low-frequency radiation component correspondingly disappears or appears \citet{rajwade2021MNRAS.506.5836R}. By separating the shifted pulses from the normal pulses, it was found that their polarization position angle (PA) curves exhibit differences. Through RVM (Rotating Vector Model) fitting and emission height calculations, it was confirmed that the two components originate from different heights \citep{sun2022ApJ...934...57S,2024_Cai}. 

The frequency-dependent radiation characteristics are
often explained based on variations in radiation heights. Hence, the model explaining the shifted pulses has many similarities to the one 
explaining the observations of PSR B$0031-07$. This may be also the most competitive model to explain the observational phenomena of PSR J$1948+3540$. However, as discussed in the previous subsection, unlike PSR B$0031-07$, the radiation component $\rm{M_{I}}$ of PSR J$1948+3540$ does not disappear at higher frequencies; rather, it ceases to dominate the intensity modulation. It is difficult to understand the variation of the modulation dominant component with observing frequency, especially considering the minimal changes in the spectra of these two components. In addition, PSR J$1948+3540$ exhibits both forward-shifted (the leading component) and backward-shifted pulses (the trailing component), and they are phase-locked. The current simple theory of magnetospheric contraction seems unable to explain the alternating occurrence of these phenomena. 

\section*{Acknowledgments}
This work is supported by the open research project funded by the Key Laboratory of Xinjiang Uyghur Autonomous Region ($2021000059$), the Natural Science Foundation of China ($12203093$), the National Key Research and Development Program ($2022YFA1603104$), the Major Science and Technology Program of Xinjiang Uygur Autonomous Region ($2022A03013-2$), the National Natural Science Foundation of China (NSFC) project (No. $12273100$, $12041303$), the West Light Foundation of Chinese Academy of Sciences (No. WLFC $2021-XBQNXZ-027$), the National Key Program for Science and Technology Research and Development and the National SKA Program of China (No. $2022YFC2205201$, $2020SKA0120200$) and the National Natural Science Foundation of Chinagrant (No. $12288102$),  
the National Science Foundation of Xinjiang Uygur Autonomous Region ($2022D01D85$), the Tianchi Talent project, and the CAS Project for Young Scientists in Basic Research ($YSBR-063$), the Tianshan talents program ($2023TSYCTD0013$), and the Chinese Academy of Sciences （CAS）“Light of West China”Program （No. $xbzg-zdsys-202410$  and No. $2022-XBQNXZ-015$)

\software{DSPSR \citep{van2011PASA...28....1V}, PSRCHIVE \citep{Hotan2004PASA...21..302H} and TEMPO2 \citep{Hobbs2006MNRAS.369..655H}}

\bibliography{sample63}{}
\bibliographystyle{aasjournal}

\end{CJK}
\end{document}